\documentclass[times,twocolumn,final]{elsarticle}

\usepackage{medima}
\usepackage{framed,multirow}
\usepackage{xcolor}
\usepackage{amssymb}
\usepackage{latexsym}
\usepackage{cancel}
\usepackage{url}
\usepackage{xcolor}

\usepackage{hyperref}
\usepackage{amsmath,amssymb,amsfonts}
\usepackage{algorithmic}
\usepackage{graphicx}
\usepackage{textcomp}
\usepackage{soul}

\usepackage{chngcntr}
\usepackage{placeins}
\usepackage{amsfonts}       
\usepackage{nicefrac} 
\usepackage{amsmath,amssymb} 
\usepackage{amstext}
\usepackage{multirow}
\usepackage{tabu}
\usepackage{svg}
\usepackage{bbm}
\usepackage{diagbox}
\usepackage[english]{babel}
\usepackage{comment}
\usepackage{array}
\usepackage{amstext}
\usepackage{makecell}
\usepackage{caption}
\usepackage{tablefootnote}
\usepackage{babel}
\usepackage{booktabs} 
\usepackage{float}
\usepackage{graphicx}
\usepackage{multirow}
\newcommand{\revisedprev}[1]{{\color{black} #1}}
\newcommand{\revisedtwo}[1]{{\color{black} #1}}

\definecolor{newcolor}{rgb}{.8,.349,.1}
\newcommand{\impulse}[1]{\emph{#1}}
\newcommand{\impulsep}[1]{\emph{#1}}

\definecolor{gold}{rgb}{0.0, 0.0, 0.0}
\definecolor{silver}{rgb}{0.0, 0.0, 0.0}

\newcommand{\eg}{\textit{e.g.}}
\newcommand{\ie}{\textit{i.e.}}

\journal{Medical Image Analysis}

\begin{document}

\verso{Kangxian Xie \textit{et~al.}}

\begin{frontmatter}



\title{Template-Guided Reconstruction of Pulmonary Segments with Neural Implicit Functions}

\author[1,2,3]{Kangxian \snm{Xie}\fnref{equal,fn1}} 

\author[4]{Yufei \snm{Zhu}\fnref{equal}} 
\author[5]{Kaiming \snm{Kuang}} 
\author[4]{Li \snm{Zhang}} 
\author[6]{Hongwei Bran \snm{Li}} 
\author[3]{Mingchen \snm{Gao}} 
\author[1,2]{Jiancheng \snm{Yang}\corref{cor1}} 

\fntext[equal]{These authors contributed equally to this work.}
\fntext[fn1]{This work was conducted during K. Xie's research internship.}
\cortext[cor1]{Corresponding author: Jiancheng Yang (jiancheng.yang@aalto.fi)}
\address[1]{ELLIS Institute Finland, Espoo 02150, Finland}
\address[2]{Aalto University, Espoo 02150, Finland}
\address[3]{Department of Computer Science and Engineering, University at Buffalo, SUNY, NY, 14260, USA}
\address[4]{Dianei Technology, Shanghai 200000, China}
\address[5]{University of California San Diego, La Jolla, CA 92093, USA}
\address[6]{Yong Loo Lin School of Medicine, National University of Singapore, 117597, Singapore}


\begin{abstract}
High-quality 3D reconstruction of pulmonary segments plays a crucial role in segmentectomy and surgical planning for the treatment of lung cancer. Due to the resolution requirement of the target reconstruction, conventional deep learning-based methods often suffer from computational resource constraints or limited granularity. Conversely, implicit modeling is favored \revisedprev{due to} its computational efficiency and continuous representation at any resolution. We propose a neural implicit function-based method to learn a 3D surface to achieve anatomy-aware, precise pulmonary segment reconstruction, represented as a shape by deforming a learnable template. Additionally, we introduce two clinically relevant evaluation metrics to comprehensively assess the quality of the reconstruction. Furthermore, to address the lack of publicly available shape datasets for benchmarking reconstruction algorithms, we developed a shape dataset named \emph{Lung3D}, which includes the 3D models of 800 labeled pulmonary segments and their corresponding airways, arteries, veins, and intersegmental veins. We demonstrate that the proposed approach outperforms existing methods, providing a new perspective for pulmonary segment reconstruction. Code and data will be available at \url{https://github.com/HINTLab/ImPulSe}.
\end{abstract}


\end{frontmatter}


\section{Introduction}
\label{sec:introduction}

\begin{figure}
    \centering
    \includegraphics[width=1\linewidth]{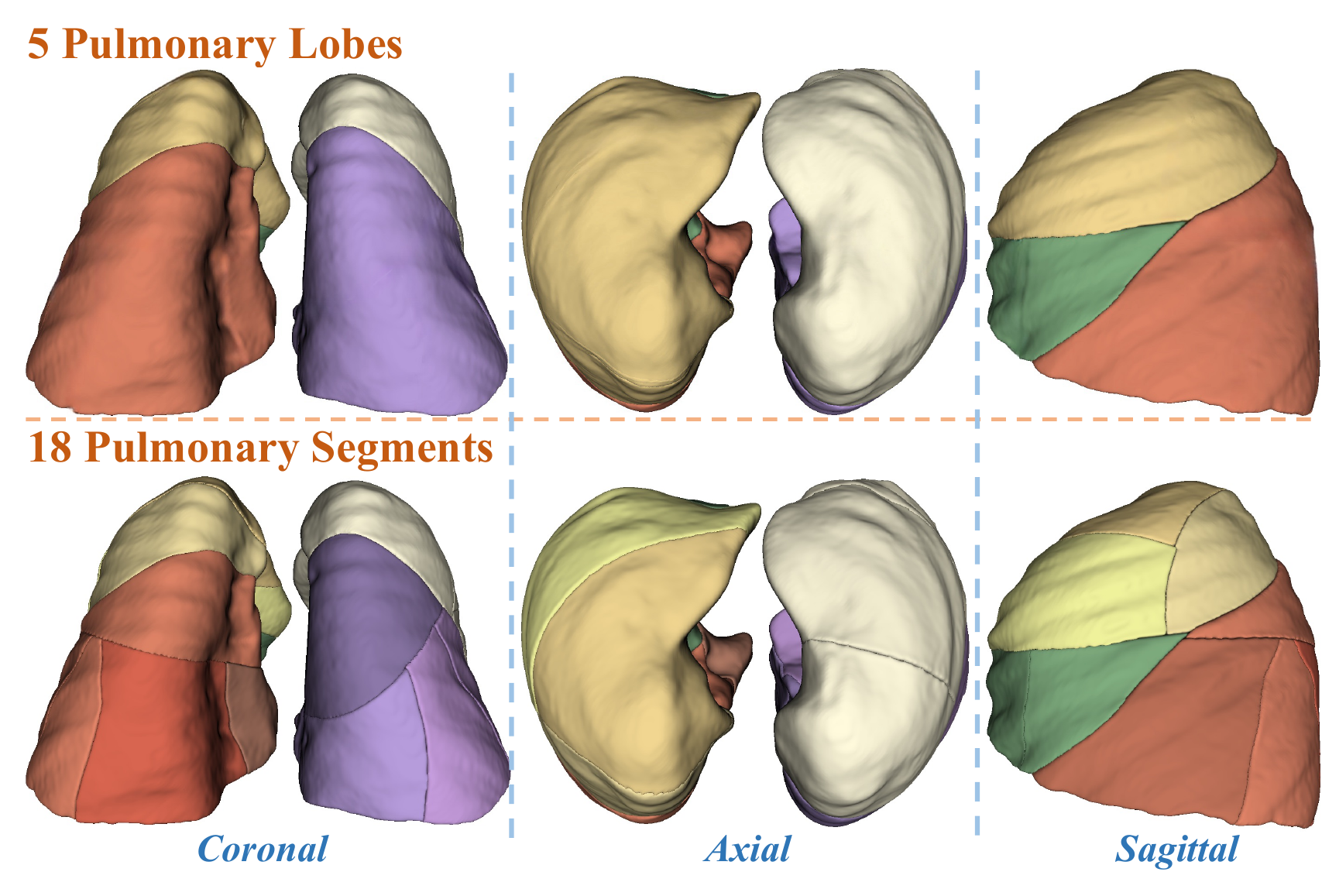}
    \caption{\textbf{A lung divided in two ways for visualization in 3 views}. The top row represents the division of the 5 pulmonary lobes, while the bottom row provides the division of the 18 pulmonary segments.}
    \label{fig:lobe_segment}
\end{figure}

\begin{figure*}[tb]
    \centering
    \includegraphics[width=\linewidth]{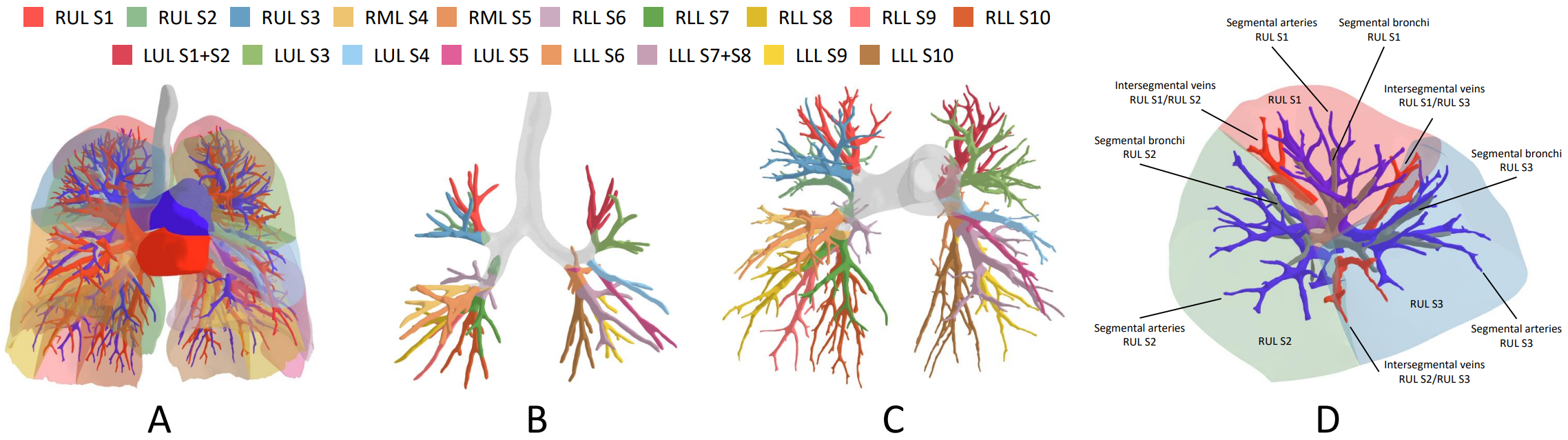}
	\caption{\textbf{Visualization of the pulmonary segment anatomy.} \textbf{A}: An overview of pulmonary segments, including bronchi, arteries, and veins. \textbf{B}, \textbf{C}: The bronchus and artery tree are divided into segmental groups, each occupying a branch of the tree. \textbf{D}: An example of intersegmental boundaries for RUL S1 (middle), RUL S2 (left), and RUL S3 (right). Segmental bronchi, segmental arteries, and intersegmental veins are colored gray, blue, and red, respectively. Intrasegmental veins are not shown for better visualization. Each segment fully encompasses its respective segmental bronchi and arteries. Intersegmental boundaries are positioned along the branches of intersegmental veins.} \label{fig:pulmonary_segment_anatomy}
\end{figure*}

Pulmonary segments are anatomically and functionally independent subdivisions of pulmonary lobes without explicit boundaries~(Fig.~\ref{fig:lobe_segment}). In lung anatomy, each segment includes its corresponding bronchus, artery, and vein, establishing its boundaries along intersegmental veins~\citep{Frick2017Segmentectomies, Oizumi2014TechniquesTD}. By anatomical definition, the right lung consists of ten segments and the left consists of eight to ten segments, depending on individual variations~\citep{Boyden1945IntrahilarRS, Jackson1943CorrelatedAA, Ugalde2007LobesFB}. The reconstruction of pulmonary segments is crucial in clinical practice, as it assists in the localization of lung diseases and the planning of surgical interventions, such as segmentectomy, a technique commonly employed for non-small-cell lung cancer because it preserves greater pulmonary function~\citep{Saji2022Jcog0802, Schuchert2007AnatomicST,Wisnivesky2010LimitedRT,Handa2021PostoperativePF,Harada2005FunctionalAR,Saji2022Jcog0802}. It would lower operation time and blood loss, resulting in lower recurrence rates and better survival outcomes. 
A crucial prerequisite for segmentectomy planning is the precise reconstruction of pulmonary segments. Traditionally, it is treated as a multi-class \textit{semantic segmentation} task in which the objective is to maximize the voxel-level prediction quality. However, we argue that the quality of pulmonary segment reconstruction does not only depend on good voxel-to-voxel matching but also on anatomical correctness~(further detail in Sec. \ref{subsec:ps_reconstruction}, and Sec. \ref{subsec:Metrics}). It is crucial that the boundaries correctly segment the class-corresponding pulmonary tree structures as shown in Fig.~\ref{fig:pulmonary_segment_anatomy}D. Hence, we term this task a \emph{reconstruction} task rather than a segmentation task.


Although deep learning-based segmentation methods have been established for pulmonary structures such as lobes, bronchi, and vessels~\citep{Gerard2019FissurenetAD, Gerard2019PulmonaryLS, Nardelli2018PulmonaryAV,zhang2023multi}, the reconstruction of pulmonary segments remains largely unexplored.
\revisedprev{CNN-based voxel-to-voxel-based methods are effective as the conventional method for performing semantic segmentation. However, in three-dimensional settings, the computational cost and memory requirement suffer from cubic growth as the input resolution increases. As the 3D CT scans are typically high-resolution, directly performing computation on the CT scans becomes impractical. Although alternative solutions may work at reduced resolution or on local patches, they produce inadequate segmentation outputs due to their limited field of view.} Since precise shape reconstruction is urgently needed for surgical navigation, the generated semantic outputs are expected to be high-resolution. This motivates us to consider alternative shape reconstruction approaches to achieve fine-grained results. 

Recent deep neural implicit functions have shown significant potential in representing continuous 3D shapes~\citep{Chen2019FLearningIF,Mescheder2019OccupancyNL,Park2019DeepSDF,Chibane2020ImplicitFF,Huang2022RepresentationSF,yang2022implicitatlas,yang2022neural,xie2023efficient,li2023medshapenet}. As they learn an implicit representation mapping coordinates to occupancy or signed distance function (SDF) at continuous locations, they are capable of reconstructing 3D shapes at arbitrary resolutions. Additionally, implicit fields can be modeled using randomly sampled points from the entire continuous space, which significantly reduces training costs. These advantages indicate that implicit functions have strong potential for reconstructing pulmonary segments.

This work extends our previous research presented at MICCAI, \impulse{ImPulSe}~\citep{Kuang2022WhatMF}, where we introduced neural implicit functions for pulmonary segment reconstruction. \revisedprev{ For this work, our four contributions could be outlined as follows. First, based on the preliminary work, we refined the problem formulation of pulmonary segment reconstruction in much better detail from an anatomical perspective. We also proposed new clinically relevant evaluation metrics to assess the quality of anatomical reconstruction.} 
Second, instead of representing a shape directly using the deep implicit surface, we enhance reconstruction quality by deforming a learned template, which integrates a template network with two implicit functions for deformation and correction. \revisedprev{Additionally, we perform analysis and visualization to further clarify and elucidate the method.} Finally, we released a shape dataset named \emph{Lung3D} (Fig.~\ref{fig:dataset})---the first benchmark for 3D reconstruction of pulmonary segments---containing the 3D models of 800 manually annotated pulmonary segments from CT images, as well as the \revisedprev{corresponding} pulmonary bronchi, arteries, and veins. On this dataset, \impulsep{ImPulSe+} enhances the performance of \impulse{ImPulSe}. 

\section{Related Works}
\subsection{Dense Prediction of Pulmonary Structures with CNNs}
\label{subsec:CNN}

Convolutional neural networks (CNNs)-based methods have been one of the most popular methods for image segmentation since the fully convolutional network (FCN) ~\citep{Long2017FCN}, DeepLabv3~\citep{Chen2017deeplabv3}, and UNet~\citep{ronneberger2015u, Isensee2021Nnunet} are established.
The recent nnUNet~\citep{Isensee2021Nnunet} emerged as a specialized UNet-based method for medical image segmentation by automatically configuring network architectures, pre-processing, and training strategies based on the input data. This data-driven approach allowed nnUNet~\citep{Isensee2021Nnunet} to achieve top performance on various medical segmentation tasks. 
These models have shown promising results in various segmentation tasks, including lung and lobe segmentation, airway segmentation, vessel segmentation, and nodule segmentation.

\revisedprev{Lung segmentation is essential for pulmonary CT analysis, as it enables the isolation of lungs for diagnostic and therapeutic applications. Various CNN-based approaches have been proposed to enhance segmentation accuracy. For instance, \cite{Khanna2020ResUNet} developed a Residual U-Net with a false-positive removal algorithm, achieving better robustness through deeper networks with residual units. \cite{Fan2020InfNetAC} introduced Inf-Net, which applies implicit reverse attention and edge-attention mechanisms for accurate infection segmentation, achieving state-of-the-art performance in COVID-19 CT image analysis. \cite{Ma2020TowardsDL} proposed a data-efficient framework incorporating semi-supervised learning and cross-domain transfer for infection segmentation, highlighting its ability to generalize across limited labeled datasets.}

Airway segmentation and detection play a vital role in respiratory disease analysis and treatment planning~\citep{zhang2023multi}. CNN-based techniques have shown promising results in extracting the airway tree from CT scans. \cite{Juarez2018AAS} proposed a fixed-stride patch-wise sliding window fashion 3D CNN, and \cite{Meng2017TSA} introduced a dynamic Volume of Interest (VOI)-based tracking method. \cite{Qin2021LearningTSCNN} proposed a 3D UNet architecture with feature calibration and an attention distillation module. The proposed method utilizes a spatial-aware feature recalibration module and a gradually reinforced attention distillation module to improve feature learning and target tubule perception.

\revisedprev{Vessel and artery detection and segmentation} are essential for surgery planning. CNN-based methods also demonstrate notable advancements in this area. \cite{Cui2019PSS} proposed a 2.5D CNN-based network (applied from three orthogonal axes for pulmonary vessel segmentation) with slice radius and multi-planar fusion, resulting in lower network complexity and memory usage compared to 3D networks. \cite{Qin2021LearningTSCNN}  introduced a novel approach for pulmonary artery segmentation via a pulmonary airway distance transform map and lung segmentation. Their proposed method demonstrates accurate segmentation of pulmonary arteries from non-contrast CT scans. \revisedprev{Additionally, \cite{Zhang2020BranchAwareDD} utilized a branch-aware CNN-based approach for artery tracking, which enforces anatomical correctness by detecting branches and radii to ensure structurally coherent reconstructions.}

Despite the advancements mentioned above, methods that generate voxel-to-voxel dense predictions face several limitations. One significant drawback is the high computational cost when applying 3D convolutions to high-resolution images. To mitigate memory constraints, these methods often operate on local patches; however, this approach compromises their ability to produce high-quality shapes and maintain consistency across local shapes. Additionally, CNN-based methods exhibit limitations in preserving geometric and topological structures. While geometric deep learning approaches, such as neural implicit functions, have shown promise, their integration into the pulmonary segment construction remains limited. 

\subsection{Neural Implicit Functions}
\label{subsec:Implicit}
Neural implicit functions have emerged as a promising avenue for shape modeling and super-resolution in various applications in medical imaging.
In shape reconstruction, \revisedprev{\cite{Chen2019FLearningIF} utilized implicit fields to enable shape extraction as an iso-surface by determining whether each point is inside or outside the 3D shape.} \cite{Park2019DeepSDF} developed DeepSDF \revisedprev{to learn} continuous signed distance functions for shape representation. Its novelty lies in the ability to map latent spaces to complex shape distributions in 3D. \cite{Mescheder2019OccupancyNL} introduced Occupancy Networks, which represent 3D reconstructions in function space, allowing for the simultaneous representation of multiple objects with high-resolution meshes. \revisedprev{\cite{Khan2022INR,Marimont2022ImplicitUF,SørensenKristine} recognized the memory requirement drawback of convolution-based models in processing high-resolution 3D medical images, and proposed neural implicit functions for reconstruction of organ and tumor with convolutional features extracted from CNN-based encoders.} \revisedprev{\cite{amiranashvili2022learning} took advantage of the continuous representation of neural implicit function to reconstruct complete 3D medical shape from sparse measurements.} \revisedprev{\cite{Raju2021DISSM} introduced deep implicit statistical shape models (DISSMs) for 3D shape delineation from medical images. DISSMs combine the strengths of deep networks and statistical shape models, employing an implicit representation to generate compact and informative deep surface embeddings, which enable statistical models of anatomical variance.} \cite{yang2022implicitatlas} developed a template-based neural implicit method producing high-quality reconstruction learned from hundreds of medical shapes. Moreover, \cite{yang2022neural} proposed to use implicit functions with image appearance as inputs to repair low-quality human annotations on 3D medical images. 

Beyond shape reconstruction, neural implicit functions have been explored in image super-resolution \citep{mcginnis2023single, wu2022arbitrary}, shape completion \citep{shen2023transdfnet}, and k-space intensity interpolation \citep{huang2023neural}. 

Existing methods demonstrate the ability of neural implicit functions to generate high-quality surfaces. However, they are sensitive to noise and require large training datasets in the context of medical imaging. Employing atlases and templates in shape modeling has been identified as a promising approach to address these drawbacks.

\subsection{Atlases and Templates}
\label{subsec:Templates}
Atlas and template-based techniques have gained significant attention and recognition in the field of biomedical image analysis because they can effectively handle the inherent noise and large variability in these images. Probabilistic atlases have become a prevalent choice for atlas-based image segmentation~\citep{Iglesias2015Multiatlas}. With the increasing use of deep learning techniques, researchers have integrated atlases into convolutional neural networks to improve segmentation  performance~\citep{Atzeni2018ProbabilisticMC, Dong2018VoxelAtlasGAN, Huo2018SpatiallyLA}. For 3D left ventricle segmentation, \cite{Dong2018VoxelAtlasGAN} introduced VoxelAtlasGAN, which employs a template to address challenges such as lower contrast, higher noise, and limited annotations.
These approaches depend on pre-computed atlases created by combining manually annotated images. 
 
Simultaneously, template-based approaches combined with implicit surfaces have also garnered attention. \cite{Deng2021DeformedIF} introduced \revisedprev{Deformed Implicit Field}, a novel implicit field-based 3D shape representation method tailored for object category shapes, which utilizes unsupervised learning to achieve dense correspondences for objects exhibiting structural variations. \cite{Zheng2021DeepIT} proposed Deep Implicit Templates (DIT), a 3D shape representation that allows for 
conditional deformations of a template implicit function in an unsupervised manner. DIT enables learning a common implicit template for a collection of shapes, establishing dense correspondences across all shapes simultaneously. 
While these methods employ implicit techniques to predict implicit deformations around a learned template, they primarily focus on large training datasets, often at the expense of data efficiency. Furthermore, they utilize multi-layer perceptron (MLP) decoders, which do not introduce spatial reductive bias like convolutional decoders~\citep{Peng2020ConvolutionalON}, and are limited to learning a single implicit template, despite the potential benefits of multiple templates.

\section{Problem Formulation}
\label{sec:problem_formul}

\subsection{Pulmonary Segment Reconstruction}
\label{subsec:ps_reconstruction}



\revisedtwo{In this study, we aim to reconstruct all 18 pulmonary segments in 3D, taking as input both the CT volume and the corresponding binary mask of the airway and vessel tree-structures.} \revisedprev{Unlike structures with fissures that are visible to human eyes~(\eg{}, pulmonary lobes, heart chambers), the boundaries of the 18 pulmonary segments are primarily determined by the corresponding pulmonary tree structures. As illustrated in Fig.~\ref{fig:pulmonary_segment_anatomy} D, each pulmonary segment encompasses its associated segmental-level branch within the pulmonary trees---bronchi, arteries, and veins. If a pulmonary segment fails to entirely enclose its corresponding pulmonary structures, we consider the segment reconstruction to be incorrect, as Figure \ref{fig:intrusion_example} presents an example of intersegment intrusion. Additionally, the boundaries between neighboring segments should be established along the intersegmental vein~\citep{Oizumi2014TechniquesTD,Frick2017Segmentectomies}.}


\revisedprev{Therefore, the challenge of pulmonary segment reconstruction lies not just in the pixel-perfect delineation of these segments but in ensuring the anatomical correctness of the reconstructed segments according to the above criteria. Given this perspective, the problem could be better characterized as \textit{reconstruction} rather than standard \textit{segmentation}.} 

\revisedprev{In the problem setup, the initial input to the algorithm will be the 3D CT images of the lung, 3D shapes: pulmonary lobes\revisedprev{~(Fig.~\ref{fig:lobe_segment})} and pulmonary tree-like binary structures~(detailed in Sec.~\ref{sec:dataset}) such as pulmonary bronchi, arteries, and veins, which are the exact structures that implicitly defines the pulmonary segments' border. The 3D shapes come from manual annotation~(Sec.~\ref{subsec:dataset_annotation}), or model prediction~(Sec.~\ref{subsec:shape_OOD}) with segmentation network~\citep{Isensee2021Nnunet}. We aim to solve the reconstruction problem with different combinations of the above-mentioned input modalities~(Sec.~\ref{sec:experiments}). As a solution, we present an 18-class semantic reconstruction algorithm that efficiently processes the given shape-based, image-based data, and adheres to anatomical constraints while being precise in voxel-to-voxel matching.}

\subsection{Evaluation Metrics}
\label{subsec:Metrics}
Reconstruction of the pulmonary segments is considered challenging because it emphasizes anatomical correctness apart from voxel-level accuracies~\citep{Kuang2022WhatMF}. For a more comprehensive evaluation, we designed several clinically relevant anatomical-level metrics to quantify the inclusion relationships between the pulmonary segments and their intra-structures. Compounded with the commonly used voxel-level metrics, we form a two-level metric system.

\paragraph{{Voxel-Level Metrics}}
These metrics focus on voxel-wise reconstruction accuracy. We include Dice score and the normalized surface Dice (NSD).
Dice score~\citep{Bernard2018DeepLT,Bilic2019TheLT,Heller2019TheSO,Menze2015TheMB} is a widely-used evaluation metric in medical image segmentation tasks, and its formulation is as follows:
        \begin{equation}
            \text{Dice}(\mathbf{Y},\hat{\mathbf{Y}})=\frac{2\lVert\mathbf{Y}\cap\hat{\mathbf{Y}} \rVert}{\lVert\mathbf{Y}\rVert+\lVert\hat{\mathbf{Y}}\rVert}
        \label{equ:dice}
        \end{equation}
where $\lVert \cdot \rVert$ is the number of elements in the set, and $\mathbf{Y}$ and $\hat{\mathbf{Y}}$ are the ground-truth and prediction. It characterizes the similarity between ground truth and prediction at the voxel level.

Compared to Dice score, normalized surface Dice~\citep{Nikolov2018DeepLT, Seidlitz2021RobustDL} (NSD) focuses on the \revisedprev{reconstruction} surface, formulated as:
        \begin{equation}
            \text{NSD}(\mathbf{S},\hat{\mathbf{S}})=\frac{\lVert\mathbf{N}^{\mathbf{S}}_{\hat{\mathbf{S}}}\rVert+\lVert\mathbf{N}_{\mathbf{S}}^{\hat{\mathbf{S}}}\rVert}{\lVert\mathbf{S}\rVert+\lVert\hat{\mathbf{S}}\rVert}
        \label{equ:nsd}
        \end{equation}
        where $\mathbf{S}$ and $\hat{\mathbf{S}}$ are the set of surface voxels in the ground truth and prediction, and $\mathbf{N}^A_B$ denotes the voxels in set $A$ that fall into the neighborhood of set $B$.

\paragraph{{Anatomy-Level Metrics}}

\begin{figure}[h!]
    \centering
    \includegraphics[width=\linewidth]{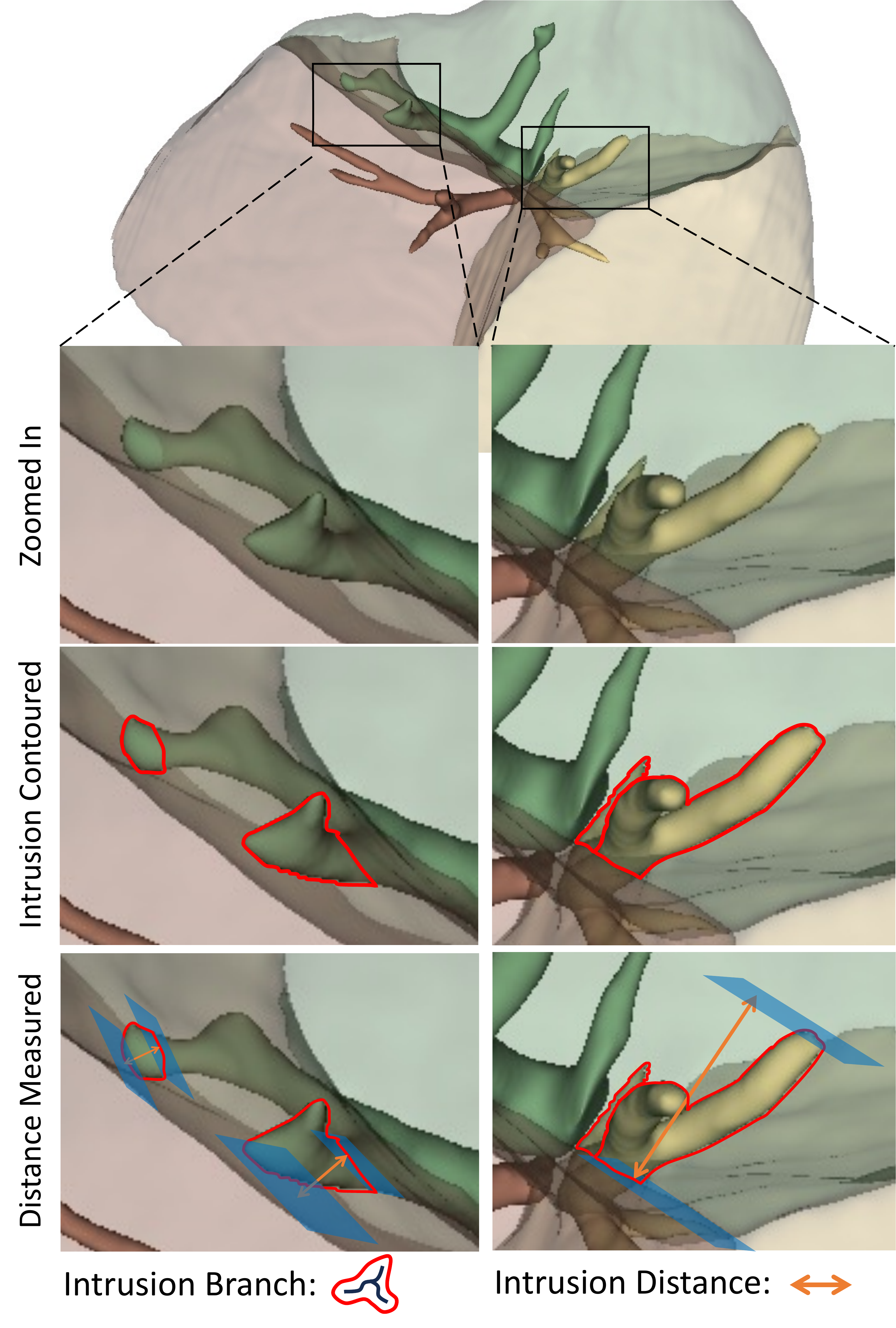}
	\caption{\textbf{Illustration of inter-segment intrusion.} Pulmonary tree structures (bronchi, arteries, and veins) intrude to neighboring pulmonary segments caused by anatomically erroneous reconstruction of pulmonary segments.} 
\label{fig:intrusion_example}
\end{figure}

These metrics reflect the high-level anatomical quality of pulmonary reconstruction. According to the anatomical definition~\citep{Kuang2022WhatMF}, pulmonary segments are defined to include their class-corresponding segmental bronchi and artery branches. Therefore, anatomically accurate segment reconstruction would avoid the intrusion of pulmonary tree branches into neighboring pulmonary segments. \revisedprev{Two Illustrations of intrusions are presented in Fig. \ref{fig:intrusion_example}. For the examples in the left column, the pulmonary structures belonging to the green class intrude on another pulmonary segment labeled red. Similarly, the example in the right column shows the pulmonary structures intruding from the yellow segment into the green segment.} To evaluate the reconstruction from an anatomical perspective, we create two new metrics: the number of intrusions and intrusion distance (Fig. \ref{fig:intrusion_example}). 

Let the collection of ground truth voxels with class $i$ for pulmonary bronchi or artery tree branches be $\textbf{T}_{i}$ and predicted pulmonary segments be $\textbf{PS}_{i}$. Then, given any collection of voxels, use $g$ as a function that breaks the voxel collection into groups of connected voxels. We define the intrusion branches (IB) as a set of voxel groups for the $i$-th pulmonary segment as:
    \begin{equation}
        \text{IB}_{i} = \{I_{1}, I_{2}, ..., I_{n}\} = g( T_{i} \cup PS_{i} - PS_{i}) 
        \label{equ:intru_branch}
    \end{equation}
thus the number of intrusions (NI) is $\sum_{i=1}^{18} \lvert \text{IB}_{i} \rvert$.

For an arbitrary intrusion branch $I$, let the intruded inter-segment surface be a set of points on the surface $S$. We measure $I$'s intrusion distance (ID) as the Euclidean distance between the furthest intrusion voxel and the surface $S$, formulated as:
    \begin{equation}
        \text{ID} = \max_{b \in I} \min_{s \in S} \lVert (b-s) \rVert
        \label{equ:intru_dist}
    \end{equation}
\revisedprev{For each subject, we will take the average ID of all of the intrusion branches and report the average ID in the performance section.}

We evaluate the NI and ID for both bronchi and arteries and report four metrics: number of intrusion bronchi (NIB), intrusion distance - bronchi (IDB), number of intrusion arteries (NIA), and intrusion distance - artery (IDA). Since the boundary between adjacent segments primarily coincides with the intersegmental vein, indicating the segmentation of the vein within pulmonary sections may not be precise. In our evaluation of NI and ID, we exclude anatomical inaccuracies pertaining to the vein.

\revisedprev{Although anatomical-level metrics are specific, novel, and highly relevant to this task, we prioritize voxel-based metrics due to their stability and comprehensive evaluation of reconstruction quality. Anatomical metrics, while insightful, are sensitive to minor boundary variations, which can cause large shifts in results. Their primary purpose is to enhance interpretability by highlighting anatomical correctness.}

\subsection{Lung3D Dataset}
\label{sec:dataset}

\subsubsection{Dataset Overview}
\label{subsec:dataset_overview}
\begin{figure}[tb]
    \centering
    \includegraphics[width=\linewidth]{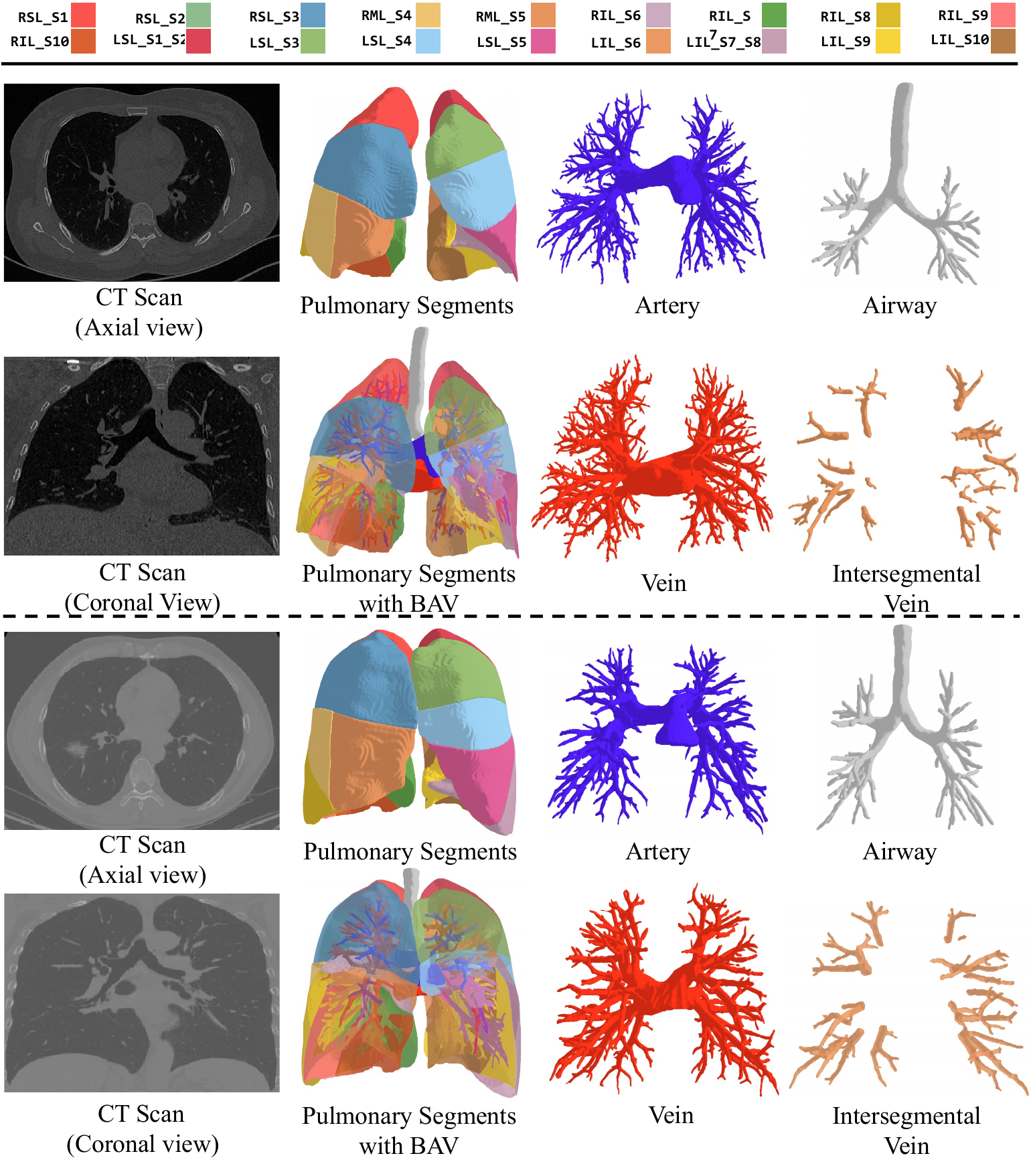}
	\caption{\textbf{\emph{Lung3D} dataset.} It consists of annotated shapes of pulmonary segments obtained from 800 multi-centered CT scans (\textbf{I}). The dataset includes annotations for pulmonary segments, bronchi/airways (\textbf{B}), arteries (\textbf{A}), veins (\textbf{V}), and intersegmental veins. The pulmonary segments are labeled with 18 classes, which can be merged into 5-class lobes (\textbf{L}).} 
\label{fig:dataset}
\end{figure}

Dependence on limited and proprietary datasets in many previous studies hinders fair and reliable benchmarking of pulmonary segment construction algorithms.  Therefore, we create a shape dataset named \emph{Lung3D}, which comprises 800 annotated cases of pulmonary segments along with associated pulmonary bronchi, arteries, and veins, \revisedprev{which include} intersegmental veins. \revisedprev{The dataset was split into 70\% training (560 subjects), 10\% validation (80 subjects), and 20\% testing (160 subjects) subsets to facilitate unbiased evaluation, respectively.} Each pulmonary segment is labeled with 18 classes. Fig.~\ref{fig:dataset} gives a visualization of our \emph{Lung3D} dataset.

\subsubsection{Data Acquisition and Annotation} \label{subsec:dataset_annotation}
\emph{Lung3D} is a multi-centered shape dataset. The original CT scans were collected from multiple public medical centers in China, including Shanghai Chest Hospital, Huadong Hospital Affiliated with Fudan University, Shanghai Pulmonary Hospital Affiliated with Tongji University, Nanfang Hospital Ganzhou, and Sun Yat-sen University Cancer Center. \revisedtwo{For each case, radiologists make manual annotations based on the CT scan. Hence, \emph{Lung3D} contains the annotated shapes of pulmonary segments (18 classes), pulmonary bronchi, arteries, and veins (binary).} 

The CT scans are stored in NIFTI (.nii) format and have volume sizes of $N\times512\times512$, where $512\times512$ represents the resolution of the CT slices, and $N$ denotes the number of CT slices, which ranges from 181 to 798. 

The annotations within the \emph{Lung3D} dataset are meticulously crafted through a collaborative process. Each annotation is labor-intensive, typically requiring approximately \textbf{3 hours} for each case. The initial annotation process is done by a junior radiologist according to the following protocol. \revisedprev{Initially, the annotations of airways were created as they serve as a prerequisite for distinguishing arteries and veins within non-contrast CT scans. Then, annotations for arteries and veins were created, with intra-segmental vein and intersegmental vein differentiated. Subsequently, the annotations of intersegmental regions were created primarily along the annotated intersegmental vein. Finally, annotations of pulmonary segments were generated according to the boundaries that were established in the previous step. This sequential approach ensured that the annotations of pulmonary segments were based on accurate delineation of airways, intersegmental regions, and other relevant anatomical structures. Finally, a senior radiologist confirms the manual annotation for accuracy and consistency. Due to the unique anatomy of pulmonary segments (Sec. \ref{subsec:ps_reconstruction}), each segment lacks a clearly defined boundary surface derived solely from image contrast. Therefore, manual labeling based exclusively on image contrast may introduce bias into the final reconstruction.} \revisedtwo{Finally, we note that the original CT scans will not be made publicly available to comply with data protection laws and safeguard patient privacy.}

\begin{figure*}[t!]
    \centering
    \includegraphics[width=\linewidth]{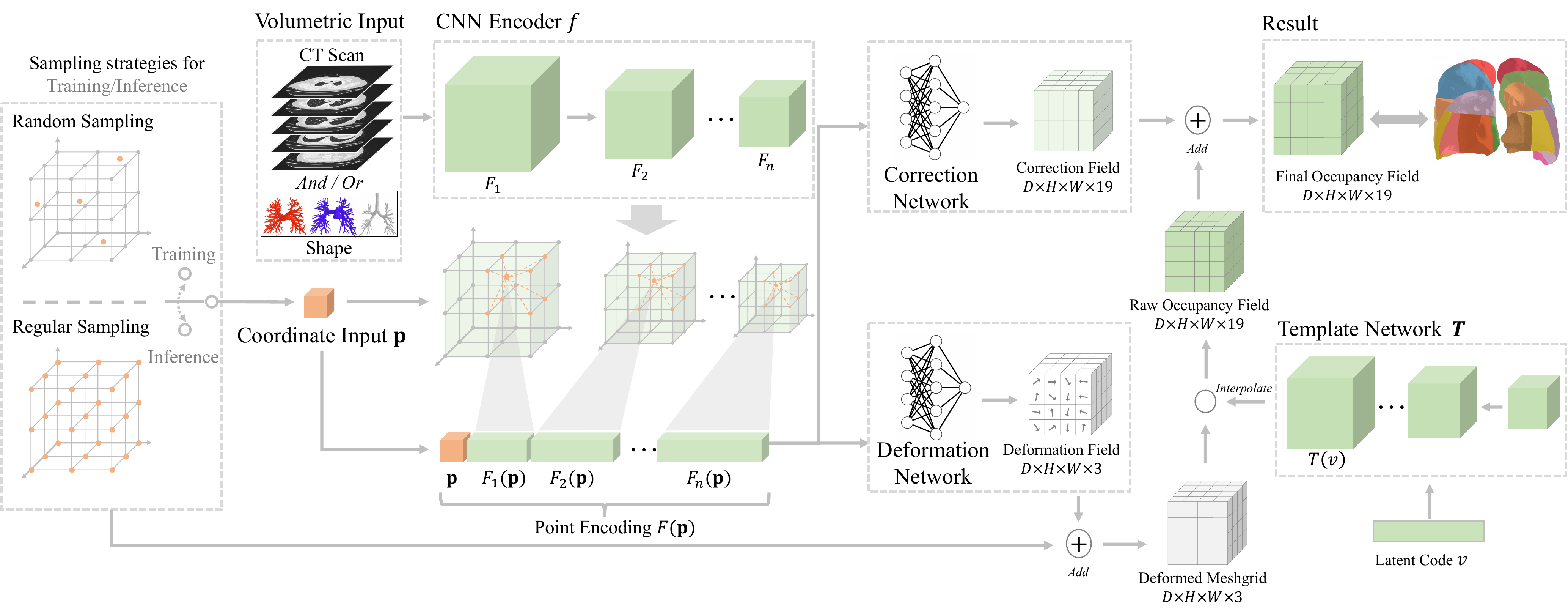}
	\caption{\textbf{Overview of \impulsep{ImPulSe+}.} The architecture includes a backbone CNN encoder $f$, a Template Network $T$, an Implicit Correction Network $C$, and an Implicit Deformation Network $D$. \revisedprev{Given the CT scan or shape input, $f$ generates the \revisedprev{Point encoding} vector $\emph{F}(\textbf{p})$} for query points, which is then fed into $D$ to predict the deformation field $d$, to align with the template $t$, and is fed into $C$ to predict the correction field $c$.} \label{fig:pipeline}
\end{figure*}

\section{Methodology}

Reconstructing pulmonary segments using conventional voxel-to-voxel dense prediction-based methods poses challenges due to the varying sizes of CT images. First, they demand substantial memory resources, especially when applied to high-resolution 3D volumes, thereby limiting their utility in high-resolution data scenarios. Second, when operating at \revisedprev{reduced} resolutions, these methods yield coarse segmentation, which is inadequate for this particular task. In contrast, implicit functions represent continuous iso-surfaces of shapes, and generate outputs at arbitrary resolutions, even with low-resolution inputs.
As a solution, we propose an implicit function-based approach that begins with a pre-trained template network and employs two implicit functions to transform and correct the \revisedprev{fixed pre-trained} template, ultimately achieving the desired reconstruction.

\subsection{Preliminaries: Neural Implicit Function}
The implicit function is typically represented as a signed distance function (SDF) or occupancy function given a query voxel grid coordinate $\textbf{p}$ as input. While SDF denotes the signed distance between a given coordinate and the nearest point on the surface of the 3D shape $S$, the occupancy function ${F}$ maps the input 3D coordinates usually to an occupancy output $z \in [0,1]$ as the probability of the point belonging to a specific class or a series of features associated with the location. Let $n$ be the output classes, the mapping can be formulated as:

\begin{equation}
{F}(\textbf{p}) = z : \mathbb(R)^3 \rightarrow \mathbb(R)^n
\end{equation}

\subsection{Architecture Overview}

\begin{figure}[h!]
    \centering
    \includegraphics[width=\linewidth]{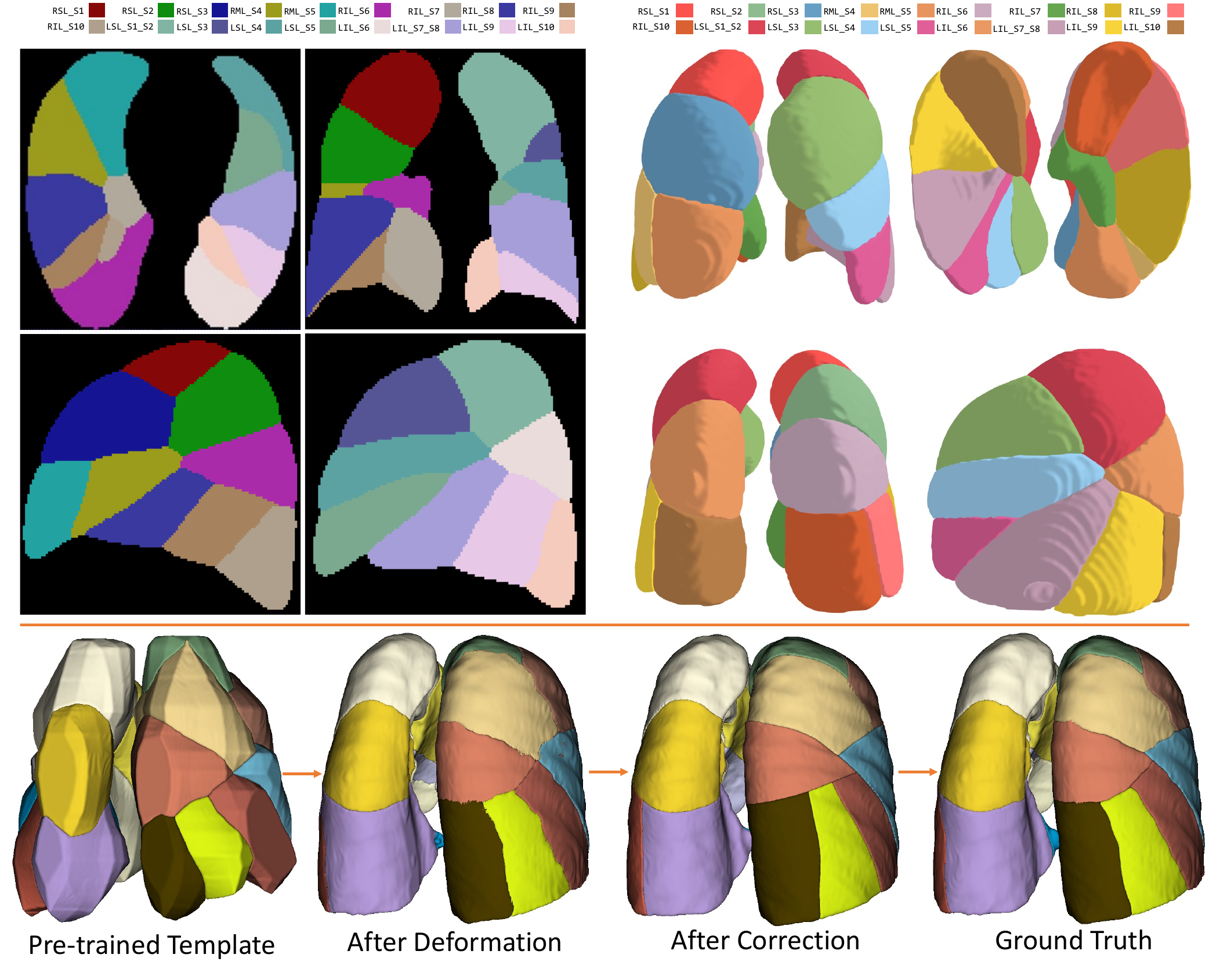}
	\caption{\textbf{Visualizations.} This figure shows (top)~the Pre-trained Template in 2D, 3D and (bottom)~the 3D visualization of pulmonary segments in each step of the architecture pipeline.} \label{fig:template}
\end{figure}

\revisedtwo{As illustrated by Fig.~\ref{fig:pipeline}, a template network ${T}$ maps a pre-trained encoding vector $\textbf{v}$ into a template of the pulmonary segment as \revisedtwo{multi-class probability distribution} $\textbf{t}$, representing the average shape of the target trainingset, illustrated in Fig.~\ref{fig:template}.

Simultaneously at the \revisedtwo{case-specific volumetric} input, the architecture takes a 3D-volume $\textbf{X}$, which can be a CT image,  binary volume of pulmonary structures (i.e., pulmonary bronchi, artery, and vein), or their combinations. Given the input, a CNN-based encoder $f$ extracts multi-scale feature pyramids $f(\textbf{X}) = \{\emph{F}_{1}, \emph{F}_{2}, ..., \emph{F}_{n}\}$. For a query point $\textbf{p}$, the corresponding multi-scale feature around point $\textbf{p}$ can be acquired through tri-linear interpolation from the feature pyramids, written as $\{\emph{F}_{1}(\textbf{p}), \emph{F}_{2}(\textbf{p}), ..., \emph{F}_{n}(\textbf{p})\}$. Subsequently, the multi-scale point feature is concatenated with the query coordinates of $\textbf{p}$ to form a point encoding $\emph{F}(\textbf{p})$ as
a feature vector that becomes the input to the implicit functions: Deformation Network D, Correction Network C. 

The Deformation Network ${D}$ and the Correction Network  ${C}$ are both implemented as multi-layer perceptrons (MLPs). Their roles are complementary to the learned template $\textbf{t}$. The Deformation Network warps and aligns the fixed template to the subject-specific anatomy. While the deformation process enforces alignment with the global template, it may not capture fine-scale variations. Hence, Correction Network refines the deformed template to account for local discrepancies. In the following sections, we will cover template generation, template deformation and final correction in detail.}

\subsection{Template Network}

The incorporation of a pre-trained template \revisedtwo{provides} the provision of prior knowledge related to probabilistic distribution and topology, as well as a reduction in the impact of noisy labels during training. Consequently, it refines the model's ability to generate outputs that exhibit more robust topological characteristics.

We introduce a template generation network ${T}$ to provide prior information for the network \revisedprev{based on the training data}. \revisedtwo{The proposed method use a fully convolutional network (FCN) decoder to implement the template network.} This network takes a parameterized latent vector $\textbf{v}$, which encodes the template, as input and outputs the pulmonary segment template $\textbf{t}$\revisedprev{~(Fig.~\ref{fig:template})}, representing the median shape of the target dataset. 

\begin{equation}
T(\textbf{v}) = \textbf{t}
\label{equ:template}
\end{equation}

Since we use an implicit function followed by the template, the resolution of $\textbf{t}$ doesn't need to be aligned with the final output resolution. Therefore, we use a shallow FCN \revisedprev{in our architecture}, \revisedprev{consisting of 8 convolutional layers,} which significantly reduces memory usage and inference time. To make sure that the common structures of the pulmonary segments are encapsulated, the latent vector $\textbf{v}$ \revisedprev{has a dimension of 1024 and} is parameterized to be trainable and randomly initialized. \revisedprev{The template network produces an output with a spatial size $128^{3}$ and 19 channels.} \revisedprev{After the template network ${T}$ and latent vector $\textbf{v}$ are fully pre-trained and incorporated with the rest of the pipeline, they are fixed and produce constant output regardless of the subject-specific volumetric input.}

\subsection{\revisedtwo{Implicit Functions: Deformation and Correction Network}}
\revisedtwo{The Deformation Network ${D}$ and the Correction Network  ${C}$ are both implemented as multi-layer perceptrons (MLPs), serving as implicit functions, that operates on point-wise features $\emph{F}(\textbf{p})$ extracted by the CNN-encoder. Their roles are complementary to the learned template. As the template represents a pre-trained and fixed multi-class shape prior, we consider any target shape as a distorted template. Thus, the Deformation Network warps and aligns the fixed template to the subject-specific anatomy. While the deformation process enforces alignment with the global template, it may not capture fine-scale variations. Hence, Correction Network refines the deformed template to account for local discrepancies. 
}

\subsubsection{\revisedtwo{Deformation Network}}

\revisedtwo{The deformation process treats the multi-class reconstruction target as a distorted version of the template, and learns to recover the necessary point-wise displacements. As denoted by Eq.\ref{equ:deformation}, for each foreground point $\textbf{p} \in \mathbb{R}^3$~(with coordinates normalized to $[-1,1]$), D takes its CNN-based point encoding $\emph{F}(\textbf{p})$ and predicts a 3-dimensional displacement vector: 
\begin{equation}
D(f(\textbf{p}), \textbf{p}) = \Delta d(p) = (\Delta x, \Delta y, \Delta z) \in [-1,1]^3
\label{equ:deformation}
\end{equation} 

$\Delta d(p)$ represents the point deviation in $x$, $y$ and $z$ direction from $p$'s original coordinate to the template space, resulting in the deformed point $\textbf{p}'=\textbf{p} + \Delta d(p)$. Applying this mapping to every point across the voxel grid yields a deformation field of size $d \times h \times w \times 3$. This field encodes how each voxel should shift in order to align with the template space.

Next, the template occupancy values $T(\textbf{v}) = \textbf{t}$, which provides the 19-class probability distribution at any spatial coordinate, is queried at the deformed positions:
\begin{equation}
\textbf{t}'(\textbf{p}) = T(\textbf{v})(\textbf{p}') \text{,  with }  \textbf{p}' = \textbf{p} + \Delta d(\textbf{p} )
\end{equation} where $\textbf{t}'$ is the deformed template. By assembling the outputs across all voxels, we obtain the raw occupancy field $\textbf{t}'$ with shape $d \times h \times w \times 19$. This field represents an initial, globally aligned reconstruction but may still exhibit local inaccuracies.}

\subsubsection{\revisedtwo{Correction Network}}
\revisedtwo{To refine this coarse reconstruction as raw occupancy field $\textbf{t}'$, we introduce the Correction Network C. For each point $\textbf{p}$, the Correction Network again takes the encoding $\emph{F}(\textbf{p})$ and predicts a 19-dimensional correction vector:
\begin{equation}
 C (\emph{F}(\textbf{p}),\textbf{p}) = \Delta c(\textbf{p}) \in \mathbb{R}^{19}
\end{equation}

The Correction network output functions as a class-wise additive adjustments to the raw occupancy field at each point location. Let the final Occupancy field be $\emph{O}$, we have the final prediction of a point as: 
\begin{equation}
 \emph{O}(\textbf{p}) = \textbf{t}'(\textbf{p}) + \Delta c(\textbf{p})
\end{equation}

In the scope of the entire 3D volume, $C$ generates a 19-class correction field with shape $d \times h \times w \times 19$. The summation between the raw occupancy field and correction field constitute the final occupancy field with shape $d \times h \times w \times 19$ for 19-class reconstruction.} 

\subsection{Training and Inference}

The training of our pipeline consists of two stages: (1)~the template pre-training, and (2)~the training of the entire pipeline. 

The template pre-training stage ensures that the parameterized vector $\textbf{v}$ encodes the overall shape distribution of the training dataset and the template network learns to generate appropriate pulmonary segments given the input. \revisedprev{To achieve this, a trainable latent vector $\textbf{v}$ is randomly initialized.} During pre-training, the template network $T$ consumes $\textbf{v}$ and generates the pulmonary segments \revisedprev{reconstruction}, which is compared against a random pulmonary segment \revisedprev{from the training set} with a weighted combination of cross-entropy loss and Dice loss. After pre-training, \revisedprev{the template network} $T$ and the latent code $\textbf{v}$ are fixed.

The second stage involves training the CNN encoder $f$, Deformation Network, and Correction Network in an end-to-end manner given a fixed template. 
During training, we employ a random sampling strategy for points $p$ $\in$ $[-1, 1]^{3}$ throughout the entire image space and this strategy offers several advantages. First, it imposes data augmentation and alleviates over-fitting, which are common issues in training deep learning models on limited datasets.  Additionally, it ensures comprehensive coverage of the entire space with fewer points. Contrary to canonical segmentation which operates on the complete voxel grid, random sampling is more efficient for training. For example, we trained \impulsep{ImPulSe+} with substantially fewer points in each batch (e.g., $16^{3}$ random points versus $64^{3}$ or $128^{3}$), and achieved superior performance. Moreover, random sampling enables greater flexibility in selecting training points and reduces the computational burden of processing large datasets. As training coordinates are continuous, the ground-truth labels are queried using nearest-neighbor interpolation. 

Similar to pre-training of the Template Network, the loss function for the \revisedprev{reconstruction} task is a weighted combination of cross-entropy\revisedprev{~(CE)} loss and Dice loss.
\begin{equation}
\mathcal {L}_{Task} = \alpha\mathcal {L}_{CE} + \beta\mathcal {L}_{Dice} \; , 
\end{equation}
\revisedprev{where $\mathcal {L}_{CE}$ has weight $\alpha = 0.5$ and $\mathcal {L}_{Dice}$ has weight $\beta=1$.} To restrict the outputs to the topology of the template prior and to mitigate Deformation Network over-fitting, we introduce a regularization penalty on the deformation field, \revisedprev{as deformation loss,}
\begin{equation}
\mathcal \revisedprev{{L}_{D}} = ||\mathbf {d}||_2 \; ,
\end{equation}
where $\left\lVert d \right\rVert_2$ denotes the L2-norm for the deformation field.

During inference, the random sampling strategy is replaced with a uniform sampling of all coordinates on the input voxel grid, so the prediction for the entire 3D volume can be provided and the output resolution is aligned with the original CT image.

\section{Image-based Reconstruction: From CTs to Segments}
\label{sec:experiments}

\subsection{Experiment Settings}
\label{subsec:exp_setting}
\revisedtwo{In the image-based experiments, we analyze the reconstruction quality when only image is available.} \revisedprev{We compare the performance of the proposed method against multiple baselines, including that of the preliminary work\citep{Kuang2022WhatMF}} with \revisedprev{ the 3D} CT images of lungs \revisedprev{being the primary input source}. As discussed in Sec.~\ref{subsec:ps_reconstruction}, pulmonary trees, including bronchi (B), arteries (A), and veins (V), are crucial for the reconstruction of pulmonary segments. Therefore, in addition to the original CT images (I), we also \revisedtwo{want to} utilize the \revisedprev{binary} pulmonary tree shapes~(Fig.~\ref{fig:dataset}: 3rd \& 4th column) as input. Therefore, we train \revisedtwo{vanilla} nnUNet~\citep{Isensee2021Nnunet} to obtain the binary pulmonary structure shapes \revisedtwo{from images} as additional input to the architecture to evaluate our approach, which we refer to as Pre-Seg shapes. \revisedtwo{Compared against ground-truth annotation, Pre-Seg training shapes have dice scores of 77.1\%, 82.9\% and 81.9\% for B,A,V respectively.} The CT image (I) and the Pre-Seg shapes are concatenated \revisedtwo{channel-wise}, referred to as {\it IBAV} input. Unless otherwise specified, {\it IBAV} serves as the default input for our image-based reconstruction. Further details on the performance differences among various input \revisedtwo{combinations} will be discussed in Sec.~\ref{sec:shape}. In the \revisedprev{image-based} experiments with {\it IBAV} input, we compare our method against a variety of CNN-based voxel-to-voxel approaches for pulmonary segment segmentation, including FCN~\citep{Long2017FCN}, DeepLabv3~\citep{Chen2017deeplabv3}, nnUNet, and also compared our method with our preliminary work \impulse{ImPulSe}\citep{Kuang2022WhatMF}. \revisedprev{To calculate the proposed anatomical-level metrics and ensure fair comparisons by evaluating all models under consistent computational resources, we repeated the experiments from \impulse{ImPulSe}. This re-evaluation yielded more fine-tuned results, which differ slightly from those reported in the original work.}


\revisedprev{Performing reconstruction at the} original resolution becomes impractical due to the computational and memory \revisedprev{requirements} associated with CNN-based models at high resolutions~($N\times512\times512$). Therefore, we \revisedprev{compromise} by down-sampling the input to dimensions of $128^{3}$ and conduct experiments with FCN, DeepLabv3, and nnUNet. For FCN and DeepLabv3 methods, ResNet-18~\citep{He2015DeepRL} is utilized as the backbone to match that of \impulse{ImPulSe} and \impulsep{ImPulSe+}. In our proposed architecture including the pre-training stage and the end-to-end modeling, inputs to the CNN networks are also downsized to $128^{3}$ to match that of CNN baselines. \revisedprev{As an alternative compromise, we experiment with giving up the global context to preserve local detail by applying a sliding-window strategy with an nnUnet.} 

\revisedprev{For all experiments, the models are trained and validated using the default training and validation datasets~(Sec.~\ref{subsec:dataset_overview}), while performance metrics are reported based on evaluations conducted on the test set. For previous work \impulse{}\citep{Kuang2022WhatMF} and the current model \impulsep{}, we conducted 5 experiment runs.} \revisedprev{Additionally}, we measure the reconstruction quality with two groups of metrics (Sec. \ref{subsec:Metrics}), including the voxel-level metrics (Dice, NSD), and the four anatomical-level metrics (NIB, IDB, NIA, IDA). \revisedtwo{We use Dice metric for model selection.} For training, we apply the AdamW optimizer to minimize the combination of cross-entropy loss and Dice loss \revisedtwo{with batch size 4}, \revisedtwo{and a cosine annealing learning rate schedule, starting with $1e^{-3}$ and decaying to a minimum of $1e^{-6}$.} The experiments are based on the implementation of PyTorch 1.11.1 and Python 3.9, on a machine with 4 NVIDIA 3090 GPUs, Intel(R) Xeon(R) CPU @ 2.20 GHz, and 128 GB memory.

\begin{table}
\centering
\caption{\textbf{Performance of \impulsep{ImPulSe+}, \impulse{ImPulSe} and CNN baselines in pulmonary segments reconstruction.} All methods are evaluated based on Dice score (\%), normalized surface dice (NSD), \revisedtwo{number of intrusions (NI) - Bronchi~(B)/Artery~(A), intrusion distance~(ID) - bronchi~(B)/Artery~(A). Most methods takes input of size $128^3$, except nnUNet~\citep{Isensee2021Nnunet,10.1007/978-3-031-72114-4_47_nnunetv2}, which operates on patch by default.}} 
\label{tab:baselines}
\resizebox{\linewidth}{!}{
\begin{tabular}{lcccccc}
\hline
Methods & Dice (\%, $\uparrow$) & NSD (\%, $\uparrow$) & NIB ($\downarrow$) & IDB ($\downarrow$) & NIA ($\downarrow$) & IDA ($\downarrow$) \\ \hline
DeepLabv3~\citep{Chen2017deeplabv3} & 81.12 & 47.29 & 26.53 & 1.52 & 48.18 & 1.08 \\
FCN~\citep{Long2017FCN} & 80.98 & 48.33 & 33.18 & \bf 0.51 & 61.71 & \bf 0.58 \\
Swin-UNETR~\citep{tang2022self_swinunetr} & 81.10 & 47.34 & 31.07 & 0.68 & 51.2 & 0.87 \\ 
Swin-UNETR v2~\citep{10.1007/978-3-031-43901-8_40_swinunetrv2} & 82.53 & 49.69 & 27.72 & 1.72 & 46.90 & 1.47 \\ 

nnUNet~\citep{Isensee2021Nnunet} & 84.58 & 61.69 & 26.30 & 2.22 & 52.36 & 1.53 \\ 
nnUNetV2~\citep{10.1007/978-3-031-72114-4_47_nnunetv2} & 84.73 & 62.01 & 49.16 & 0.80 & 122.06 & 0.65 \\

\hline
{\color[HTML]{34696D} \textit{Neural Implicit Functions }}\\
\impulse{ImPulSe}~\citep{Kuang2022WhatMF} & 85.31~($\pm$ 0.05) & 60.22~($\pm$ 0.27) & \bf 24.09 & 2.10 & 43.73 & 1.27 \\
\impulsep{ImPulSe+} & \bf 86.06~($\pm$ 0.05) & \bf 62.75~($\pm$ 0.07) & 24.21 & 2.67 & \bf 43.29 & 1.33 \\ \hline
\end{tabular}}
\label{tab:main_table}
\end{table}

\subsection{Comparative Performance Analysis}
\label{subsec:perf_analy}
Tab.~\ref{tab:main_table} shows the performance of all metrics compared with existing methods. \revisedtwo{Both iterations of nnUNets~\citep{Isensee2021Nnunet, 10.1007/978-3-031-72114-4_47_nnunetv2} utilizing sliding-window on high-resolution input achieves great results in voxel-level}, likely due to the detailed surface boundary from original resolution. In comparison, the experiments taking global yet low-resolution ($128^{3}$) inputs produce inferior voxel-level metrics, \revisedtwo{such as both versions of the Swin-UNETR~\citep{tang2022self_swinunetr,10.1007/978-3-031-43901-8_40_swinunetrv2},} due to information loss after downsizing, a necessary procedure for practicality. The two implicit-based methods allow for direct generation of reconstruction at the original resolution, thereby yielding results with fine-grained details, and achieving high voxel-level performance. Between them, the proposed \impulsep{ImPulSe+} demonstrates superior performance, achieving 0.9\% improvement in terms of Dice and 4\% improvement in NSD over the previous SOTA \impulse{ImPulSe}~\citep{Kuang2022WhatMF}, showcasing the capability for accurate surface reconstruction of the template method. \revisedprev{Both \impulse{ImPulSe}~\citep{Kuang2022WhatMF} and \impulsep{ImPulSe+} were evaluated over five runs to account for performance variability. Tab.~\ref{tab:main_table} includes standard deviations for Dice and NSD, and paired t-tests reveal p-values of $0.018$ (Dice) and $0.056$ (NSD), proving the robustness and improved performance of the proposed method over the baseline.}

\begin{figure*}
    \centering
    \includegraphics[width=\linewidth]{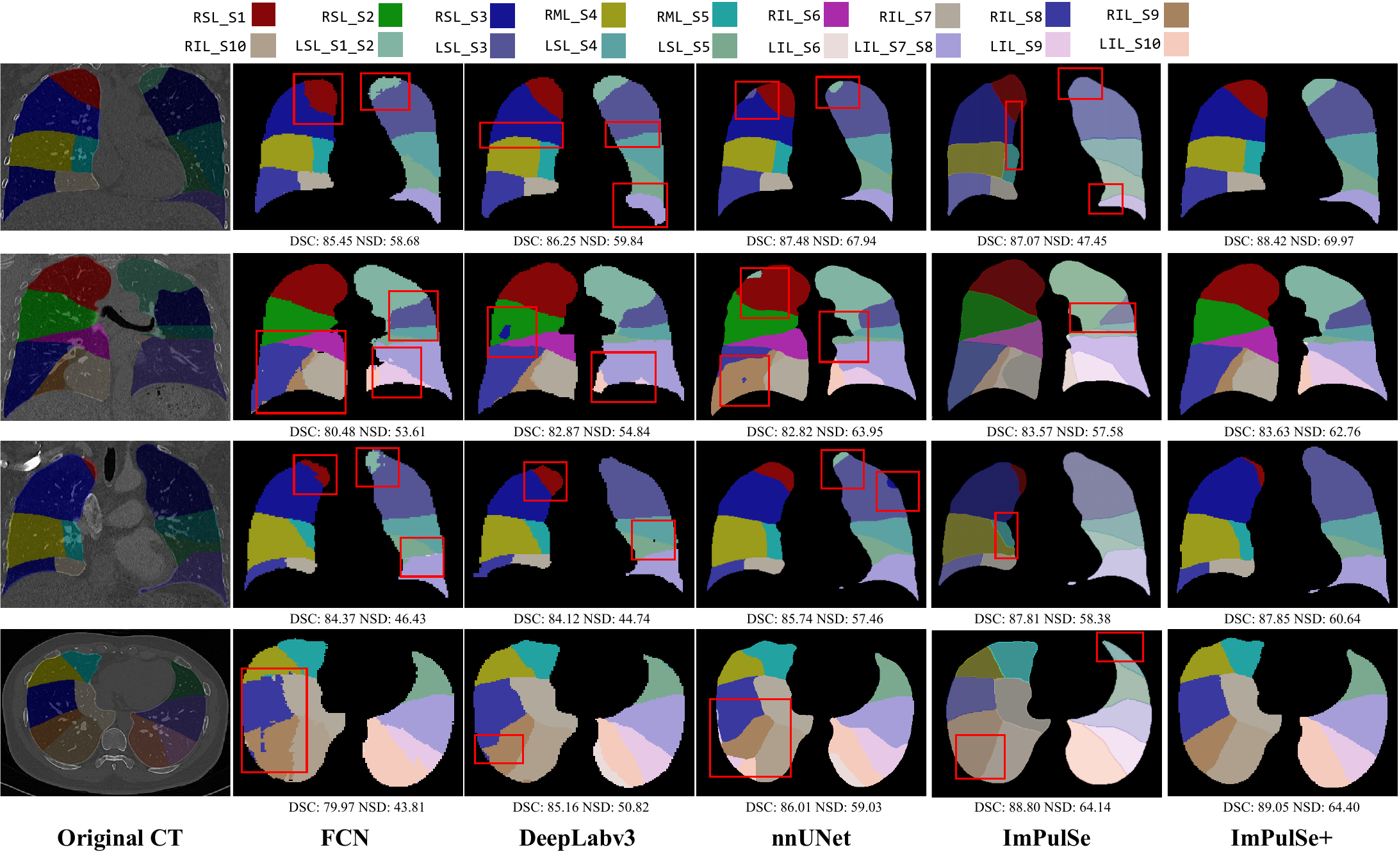}
	\caption{\textbf{Qualitative comparison in 2D visualization.} Results of the FCN, DeepLabv3, nnUNet, and our proposed neural implicit function-based method.} 
\label{fig:qualitative2d}
\end{figure*}

\begin{figure*}
    \centering
    \includegraphics[width=\linewidth]{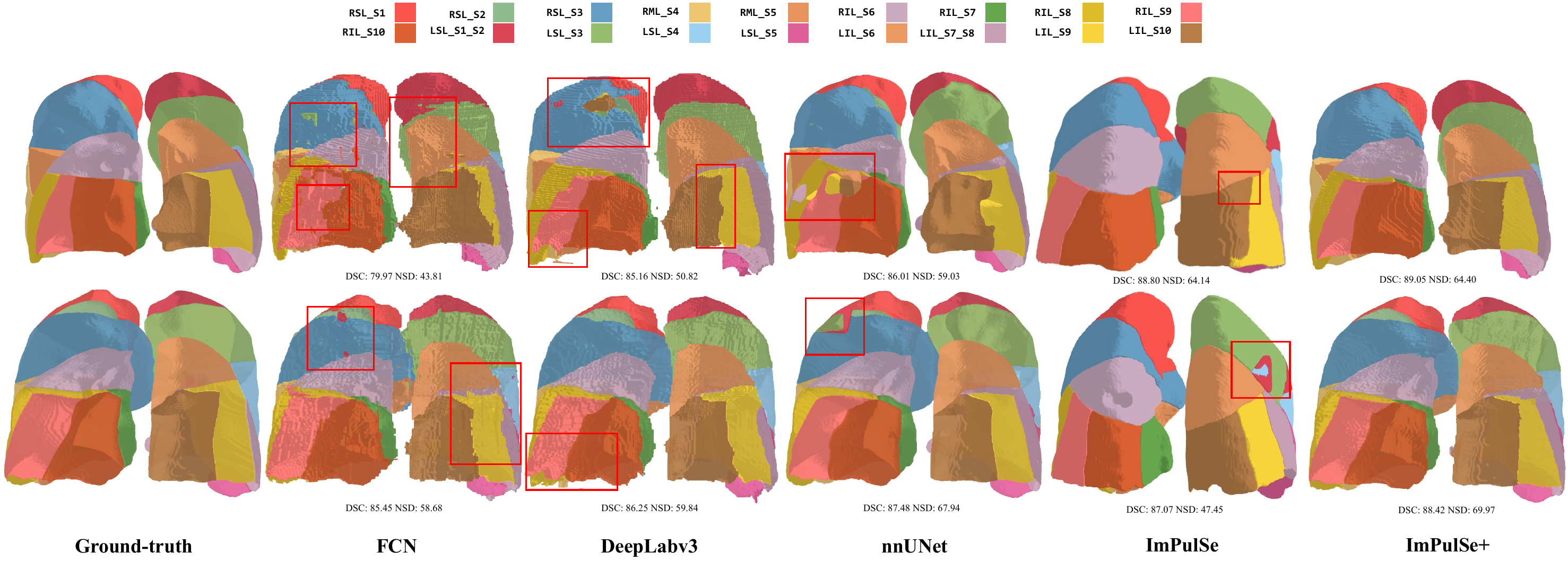}
	\caption{\textbf{Qualitative comparison in 3D visualization.} By adapting to implicit functions, the \impulsep{ImPulSe+} model is capable of directly generating segmentation results in their original size, leading to predictions characterized by smoother surfaces and superior normalized surface distance (NSD) compared to the baselines.} 
 \label{fig:qualitative_analysis}
\end{figure*}

At the anatomical level, as shown in Tab.~\ref{tab:main_table}, FCN exhibits strong performance in reducing average intrusion distances~(ID). However, this is accompanied by a significantly higher number of intrusion occurrences~(NI), resulting in compromised anatomical correctness. Similarly, while nnUNet achieves fewer intrusion occurrences, it suffers from significantly larger intrusion distances, reflecting weaker anatomical alignment. The anatomical inconsistency in FCN and nnUnet is also reflected in the visualizations (Fig.~\ref{fig:qualitative2d}, ~\ref{fig:qualitative_analysis}), exhibiting increased noise, which limits their applicability in clinical settings and undermines their credibility in the eyes of healthcare providers. In contrast, \impulsep{ImPulSe+} achieves a competitive, and stable performance with modest variation in all of the anatomical-level metrics. \revisedprev{Finally, It is important to note that these anatomical metrics are primarily intended to offer better interpretability of the problem rather than serve as definitive measures of overall reconstruction quality. }

\revisedtwo{In the resource consumption aspect, although the nnUNet variants~\citep{Isensee2021Nnunet, 10.1007/978-3-031-72114-4_47_nnunetv2} achieve the second best performance while requiring minimal memory usage~(\~7GB), the inference time are high due to the inherent patch-based operation, averaging to 42 seconds per reconstruction on the proposed dataset~\ref{sec:dataset}. With the proposed neural implicit-based methods~\citep{Kuang2022WhatMF}, the inference cost for full resolution 3D reconstruction averages at 6.7 seconds and \~13GB.

Considering all of the above, \impulsep{ImPulSe+} remains the best all-around method by achieving the strongest accuracy while offering a more favorable accuracy–efficiency trade-off. Compared to strong nnU-Net variants, our neural-implicit approach delivers full-resolution reconstructions with notably faster inference at moderate memory cost.}

\begin{figure}[h!]
    \centering
    \includegraphics[width=\linewidth]{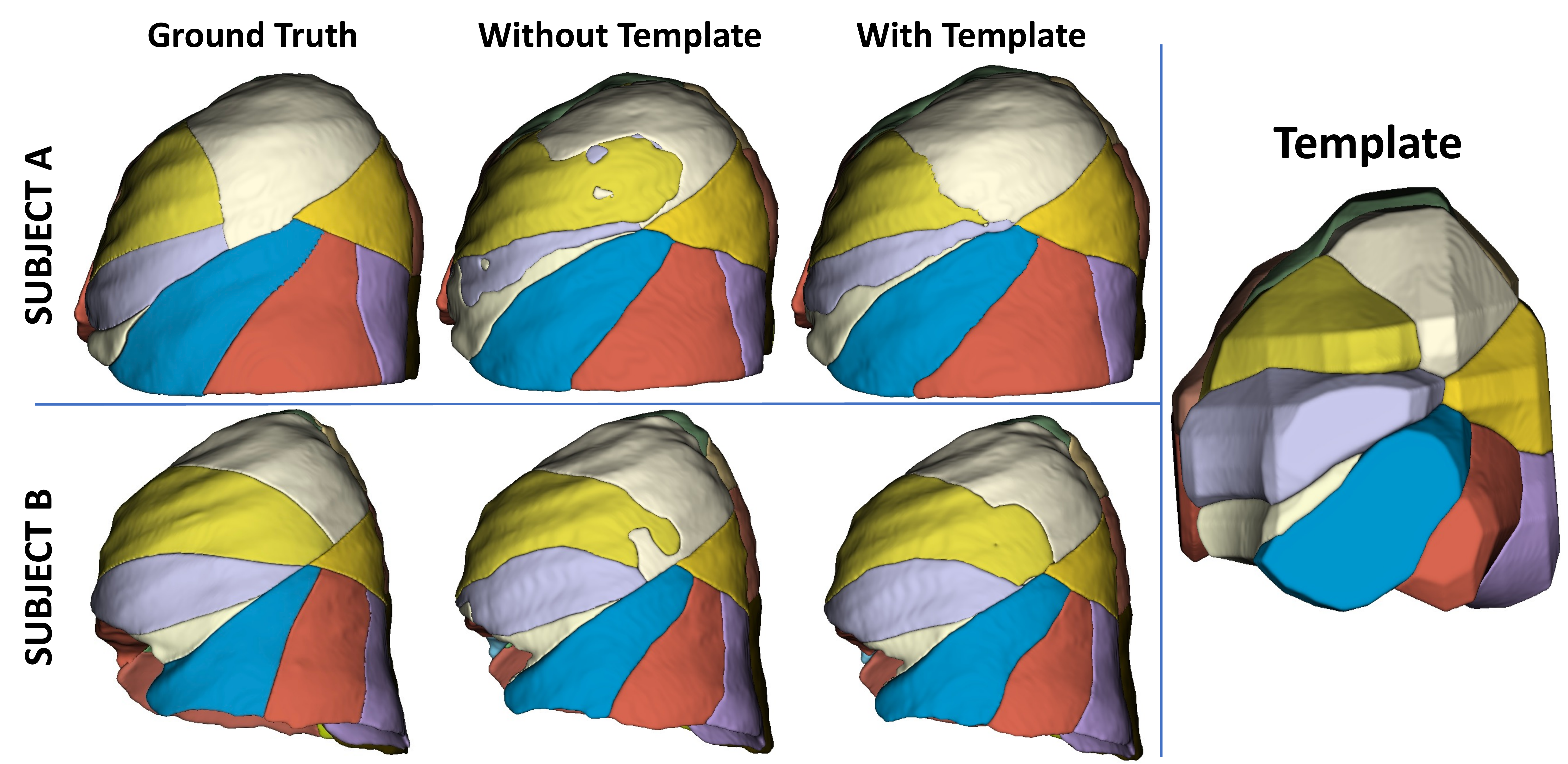}
	\caption{\textbf{The effect of template-based method.} This figure illustrate the advantage of incorporating template-based approach into neural implicit-based method.} \label{fig:template_proof}
\end{figure}

\subsection{Visualization}

\subsubsection{Qualitative Analysis}
\label{subsec:qualitative}

\revisedprev{To demonstration the high-resolution and precise reconstruction that could be achieved by \impulsep{ImPulSe+}, we conducted a qualitative analysis of the automatic reconstruction results on pulmonary segments.} Selected examples \revisedprev{by row} are presented in Fig. \ref{fig:qualitative2d} in 2D and Fig. \ref{fig:qualitative_analysis} in 3D, comparing results from FCN, DeepLabv3, nnUNet, and \impulsep{ImPulSe+} against the ground-truth.

In both 2D and 3D visualizations, results from FCN exhibit poor reconstruction outcomes, characterized by noisy and incorrect pulmonary segments, indicated by red boxes. Compared to FCN, DeepLabv3, and nnUNet demonstrate smoother boundaries with less noise and overall better similarity against ground truth. Although they provide surfaces with higher quality, there are still instances where voxels intrude into neighboring classes as well as incorrect boundary shapes. Furthermore, in both FCN and DeepLabv3 where the input dimension is restricted, the outputs are presented in limited resolution, displaying a coarse surface and lack of fine-grain detail, especially in 3D. In contrast, our proposed method \impulsep{ImPulSe+}, powered by implicit modeling, achieves visually refined \revisedprev{reconstruction} results with smooth surfaces and sophisticated detail by directly generating segmentation at the original dimension, leading to high NSD scores. Additionally, the predicted segment boundaries are overall more accurate with minimal noise. 
\begin{figure}
    \centering
    \includegraphics[width=1\linewidth]{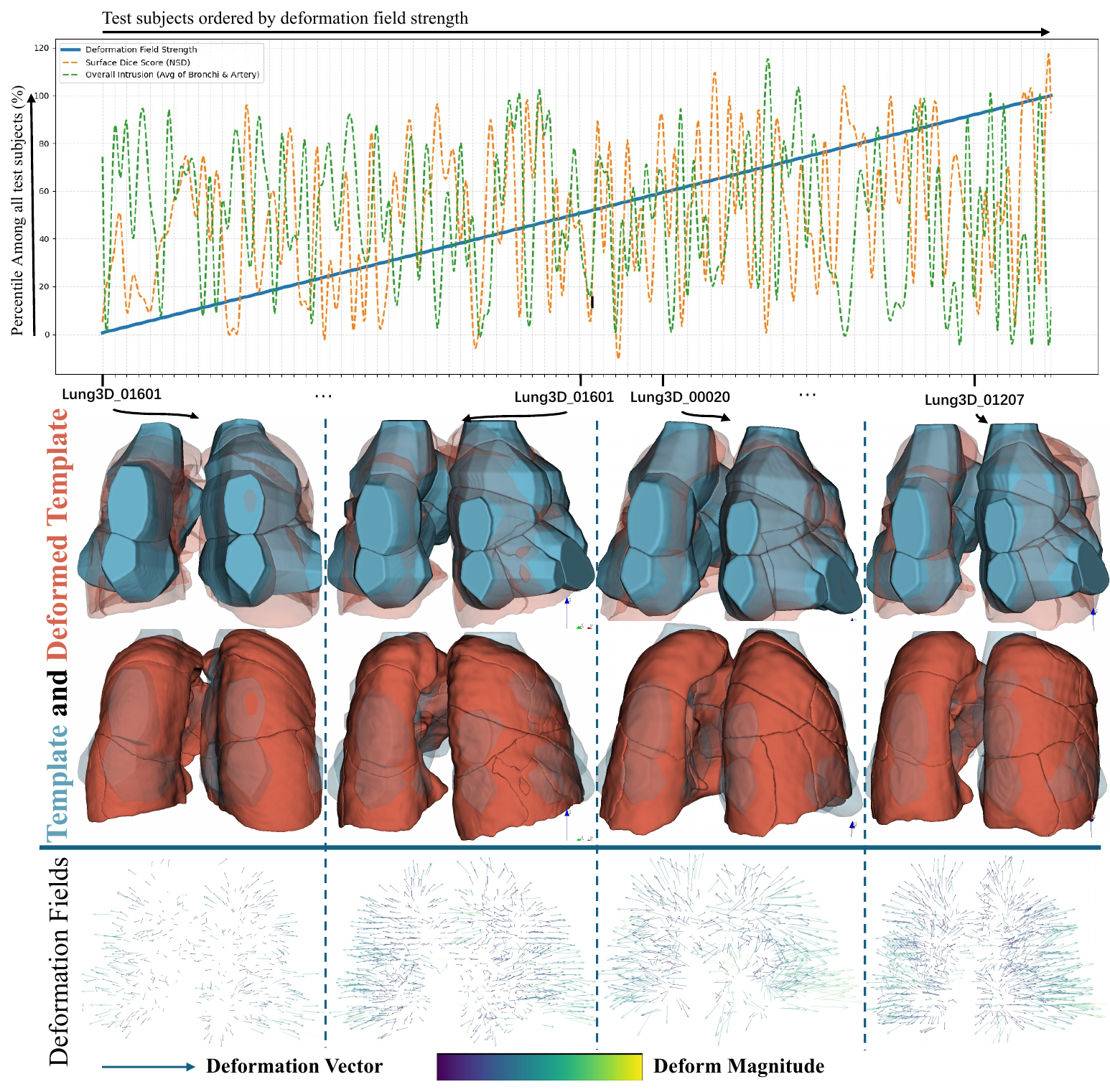}
    \caption{Sensitivity of model performance to rare anatomical variations. The plot shows deformation field strength against performance metrics. Examples with varying deformation strengths are illustrated. }
    \label{fig:correlation}
\end{figure}

\subsubsection{Pre-trained Template}
\label{subsec:template_visualization}

Before training the \impulsep{ImPulSe+} network, a pre-training phase is conducted for the template network $T$ and latent vector $\textbf{v}$. After pre-training, the generated template $\textbf{t}$, representing the mean shape of target structures, demonstrated satisfactory results, reaching $60.28\%$ in Dice score. In Fig. \ref{fig:template}, the visualization of the pre-trained template $t$ is presented in both 2D and 3D, both displaying the inter-connected components of pulmonary segments with their intrinsic topology, although not explicitly enforced. As illustrated, neighboring components in the templates are tightly connected by smooth borders, providing a clear and concrete idea of the general distribution of pulmonary segments while enabling the subsequent shape deformation. As implicit networks are learned after the template generation, the generated template $T(\textbf{v})$ can be at a lower resolution of $128^{3}$, not necessarily aligned with the final output resolution.

\revisedprev{Fig. \ref{fig:template_proof} compares the reconstruction result based on neural implicit function only~(\impulse{ImPulSe}~\citep{Kuang2022WhatMF}) against that of the proposed template-based neural implicit method. When the template is incorporated, the predictions are significantly more regularized, with smoother boundaries and less noise. These results align closely with the natural anatomical shapes of pulmonary segments. In contrast, the predictions without the template appear noisier and exhibit irregular boundary shape. This comparison illustrates the template's role in getting anatomically accurate results.}
\revisedtwo{

\subsection{Robustness and Failure Analysis}
\subsubsection{Sensitivity to Rare Anatomical Variations}
\label{subsec:sen_variation}

To assess whether our use of a fixed template introduces bias against rare anatomical cases, we analyzed model performance with respect to deformation strength, defined as the average magnitude of the top 5\% of vectors in the deformation field (Fig.~\ref{fig:correlation}, bottom). Since the template represents the mean anatomy of the training distribution, larger deformation strengths indicate greater deviations from this average and thus rarer anatomical variations. If the method were subject to template bias, where the fixed template acts as an overly strong prior, it could overfit to common anatomies~(less deformation strength) while under-representing outliers~(greater deformation strength), thereby suppressing anatomical variability.

To investigate this, we plot the deformation strength of all test subjects in ascending order~(in percentile) and overlaid the corresponding performance metrics. As shown in Fig.~\ref{fig:correlation}, there is no clear trend of decreasing Dice score or increasing anatomical-level metrics as deformation strength increases. We further quantified this relationship using Pearson correlation tests, presented in Tab.~\ref{tab:deformation_correlation}.
\begin{table}[ht]
\centering
\caption{Pearson correlation between \textbf{deformation strength} and segmentation performance metrics.}
\label{tab:deformation_correlation}
\begin{tabular}{lcc}
\hline
\textbf{Metric} & \textbf{Pearson $r$} & \textbf{$p$-value} \\
\hline
Dice score              & 0.0201  & 0.801 \\
Anat. Metrics & 0.0176  & 0.826 \\
\hline
\end{tabular}
\end{table}

 However, based on the analysis in Tab.~\ref{tab:deformation_correlation}, these results shows a non-negative correlation coefficient, and high \textit{p}-value, demonstrating weak to no correlation between deformation strength and reconstruction quality. Importantly, neither Dice nor NI/ID degrade as deformation strength increases, suggesting that stronger template deformations (i.e., rarer anatomical variations) do not compromise voxel-level or anatomy-level correctness. By observing the comparison between the template and the deformed shapes across four qualitative examples (cases $00890$, $01601$, $00020$, and $01207$), which respectively lie at the $1st$, $54th$, $62nd$, and $90th$ percentiles of deformation strength, we find that the model consistently produces anatomically plausible reconstructions even under large deviations from the template. These results illustrate that stronger deformation strength does not compromise reconstruction quality,
}
\begin{figure}
    \centering
    \includegraphics[width=1\linewidth]{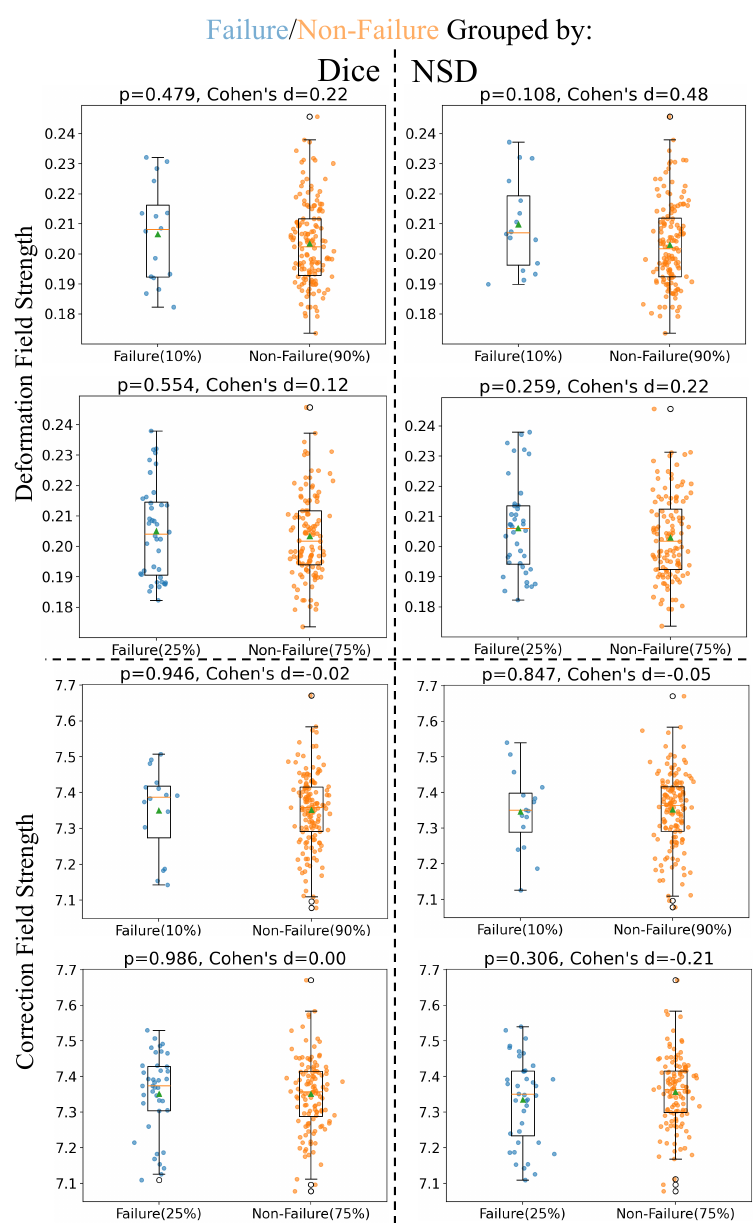}
    \caption{Failure Case Analysis. Group comparison of deformation and correction field strengths between failure and non-failure cases. Failures were defined as the bottom 10\% or 25\% of subjects according to either Dice or NSD. Each subplot shows the distribution of deformation/correction values across the two groups, with boxplots and individual subject points overlaid. Welch’s t-test p-value and Cohen's d are presented in the botplot}
    \label{fig:failurecase_boxplot}
\end{figure}
\subsubsection{\revisedtwo{Failure Case Analysis}}
\label{subsec:failurecase}
\revisedtwo{This analysis studies whether failure reconstruction cases exhibit statistically different deformation or correction strengths relative to non-failure cases. Failures were defined as the worst-performing subjects, which are the bottom 10\% or 25\% percentile test-case in terms of Dice or NSD metrics.

As defined in Sec.~\ref{subsec:sen_variation}, deformation field strength quantifies the magnitude of top 5\% deformation vectors. In addition, we define correction strength as the mean absolute value of the predicted correction field over its full spatial extent $d \times h \times w \times 19$, representing the overall correction magnitude.

The failure and non-failure cases were put into two groups for statistical analysis. According to results shown in Fig.~\ref{fig:failurecase_boxplot}, group comparisons using Welch’s t-test revealed no statistically significant differences in either measure across all definitions of failure (p$>0.05$). For deformation strength, effect sizes were generally small, with the exception of the NSD-defined 10\% failure group, which showed a moderate but non-significant trend toward higher values (d = 0.48, p = 0.108). All other comparisons yielded negligible effects (e.g., Dice-25\%: d = 0.12, p = 0.554). Correction strength demonstrated even weaker associations, with effect sizes close to zero or slightly negative (e.g., Dice-10\%: d = -0.02, p = 0.946; NSD-25\%: d = -0.21, p = 0.306), indicating no consistent pattern.

Overall, these findings suggest that neither deformation strength nor correction strength serves as a reliable discriminator between failure and non-failure cases. The borderline effect observed for deformation strength under the most stringent NSD threshold may warrant further investigation, but the absence of consistent significance or effect sizes across conditions indicates limited evidence for a direct relationship.}

\subsection{Ablation Studies}
\label{subsec:exp_Abl}
\begin{table}[ht]
\centering
\caption{\textbf{Ablation on \impulsep{ImPulSe+} network architecture design}. T: Template network. D: Deformation network. C: Correction network. $\mathcal{L}_{D}$: deformation loss. \revisedtwo{Pre. Tr.: pre-trained template network and template latent code. Co-Tr.: co-trained template with the rest of the pipeline.}}
\label{tab:ablation_experiment}
\resizebox{\linewidth}{!}{
\begin{tabular}{cccccc|cc|cccc}
\hline
\multicolumn{6}{c|}{Methods} 
& \multicolumn{2}{c|}{Voxel-level ($\uparrow$)} 
& \multicolumn{4}{c}{Anatomical-level ($\downarrow$)} \\
\cline{1-6}\cline{7-12}
\multicolumn{1}{c|}{T} & \multicolumn{1}{c|}{D} & \multicolumn{1}{c|}{C} & \multicolumn{1}{c|}{$\mathcal{L}_{D}$} & Pre.Tr. & Co Tr. 
& Dice & NSD & NIB & IDB & NIA & IDA \\ \hline
\multicolumn{1}{c|}{-} & \multicolumn{1}{c|}{-} & \multicolumn{1}{c|}{-} & \multicolumn{1}{c|}{-} & - & - 
& 85.31 & 60.22 & 24.09 & 2.10 & 43.73 & 1.27 \\
\multicolumn{1}{c|}{\checkmark} & \multicolumn{1}{c|}{\checkmark} & \multicolumn{1}{c|}{-} & \multicolumn{1}{c|}{-} & - & - 
& 84.60 & 59.92 & 29.24 & \textbf{0.62} & 44.33 & \textbf{0.90} \\
\multicolumn{1}{c|}{\checkmark} & \multicolumn{1}{c|}{\checkmark} & \multicolumn{1}{c|}{-} & \multicolumn{1}{c|}{-} & \checkmark & - 
& 84.61 & 60.35 & \textbf{23.79} & 2.23 & 48.08 & 1.41 \\
\multicolumn{1}{c|}{\checkmark} & \multicolumn{1}{c|}{\checkmark} & \multicolumn{1}{c|}{\checkmark} & \multicolumn{1}{c|}{-} & \checkmark & - 
& \textbf{86.06} & 62.75 & 24.21 & 2.67 & 43.29 & 1.33 \\
\multicolumn{1}{c|}{\checkmark} & \multicolumn{1}{c|}{\checkmark} & \multicolumn{1}{c|}{\checkmark} & \multicolumn{1}{c|}{\checkmark} & \checkmark & - 
& 86.00 & \textbf{63.17} & 24.09 & 3.29 & \textbf{42.39} & 1.43 \\

\multicolumn{1}{c|}{\checkmark} & \multicolumn{1}{c|}{\checkmark} & \multicolumn{1}{c|}{\checkmark} & \multicolumn{1}{c|}{\checkmark} & - & \checkmark
& 85.26 & 57.07 & 23.10 & 3.07 & 45.69 & 1.56 \\ \hline
\end{tabular}}
\label{tab:impulse_ablation1}
\end{table}

\subsubsection{Network Architecture Design}
\label{subsubsec:arch_Abl}
In the ablation study of the proposed \impulsep{ImPulSe+} pipeline, we aim to evaluate the contribution of its various components to the overall performance. \revisedprev{In Tab.~\ref{tab:impulse_ablation1}, we use check marks to signal if a network component is applied. The first row represents the performance of our preliminary work~\citep{Kuang2022WhatMF}. \revisedtwo{Unlike current architecture, which leverages a fixed template that is subsequently deformed and corrected, the ImPulSe architecture employs an implicit network to directly generate the 19-class reconstruction without using a template. Thus, the first-row entry in the ablation table corresponds to the “without T” setting.} The second row shows the results with non-pretrained template network co-trained with the deformation network. In the third row, the template network is pre-trained. The fourth and final rows incrementally add the Correction network and incorporate deformation loss, respectively.}

As the results indicate, when we leverage only the Template Network, either with or without pre-training, the overall performance represented by dice metrics is unsatisfactory and worse than our prior work \impulse{ImPulSe}~\citep{Kuang2022WhatMF}. After the Correction Network is integrated with the pre-trained Template Network, we achieve \revisedprev{0.9\%} dice and 4.2\% NSD performance improvements over the predecessor. Finally, as we employ the deformation loss during training, the model outperforms that without deformation loss by \revisedprev{approximately 0.7\%} in NSD, leading to a smoother and more accurate surface and comes without considerable sacrifice in Dice.

\revisedtwo{
In addition, we examined the effect of co-training the template network with the entire pipeline instead of pretraining. The results in the last row of Tab.~\ref{tab:ablation_experiment} indicate that this setting achieves voxel and anatomical-level performance comparable to the proposed architecture. However, NSD decreases notably to 57\%, suggesting less accurate surface alignment. We attribute this to the fact that updating the template during training reduces its role as a stable global anatomical prior, leading the model to rely more on local deformations at the expense of consistent boundary accuracy. This observation supports our design choice of using a fixed pretrained template, which provides a better balance between reconstruction accuracy and surface consistency.

Overall, these results demonstrate that the Template and Correction networks are critical for manipulating the learned template, achieving high quality pulmonary segment reconstructions, while the deformation loss further refines boundary alignment, leading to higher NSD without sacrificing Dice.
}

\subsubsection{Foreground Point Sampling in Training}
\label{subsec:exp_BAV}

Within this section, we conduct experiments to investigate the impact of a special sampling strategy, {\it{BAV}} sampling, during the training for the proposed \impulsep{ImPulSe+} pipeline.

In the default sampling strategy, where points are sampled across the vast 3D space, the proportion of points originating from bronchial, arterial, and venous regions is limited. To ameliorate issues concerning the intrusions of segmental bronchi and arteries into neighboring pulmonary segments, we experiment with a deliberate augmentation in the proportion of points sampled from the areas of bronchi ({\it{B}}), arteries ({\it{A}}), and veins ({\it{V}}) during training.

In this experiment, let $\gamma$ refer to the proportions of points that are randomly sampled over the entire CT space, then '1 - $\gamma$' designates the proportion of points from the {\it{BAV}} space and we test the value from 20\% to 90\%. To showcase the advantage of this strategy comprehensively, we present the results with the product of the number of intrusions and intrusion distance, as total intrusion distance. 

\begin{figure}
    \centering
    \includegraphics[width=\linewidth]{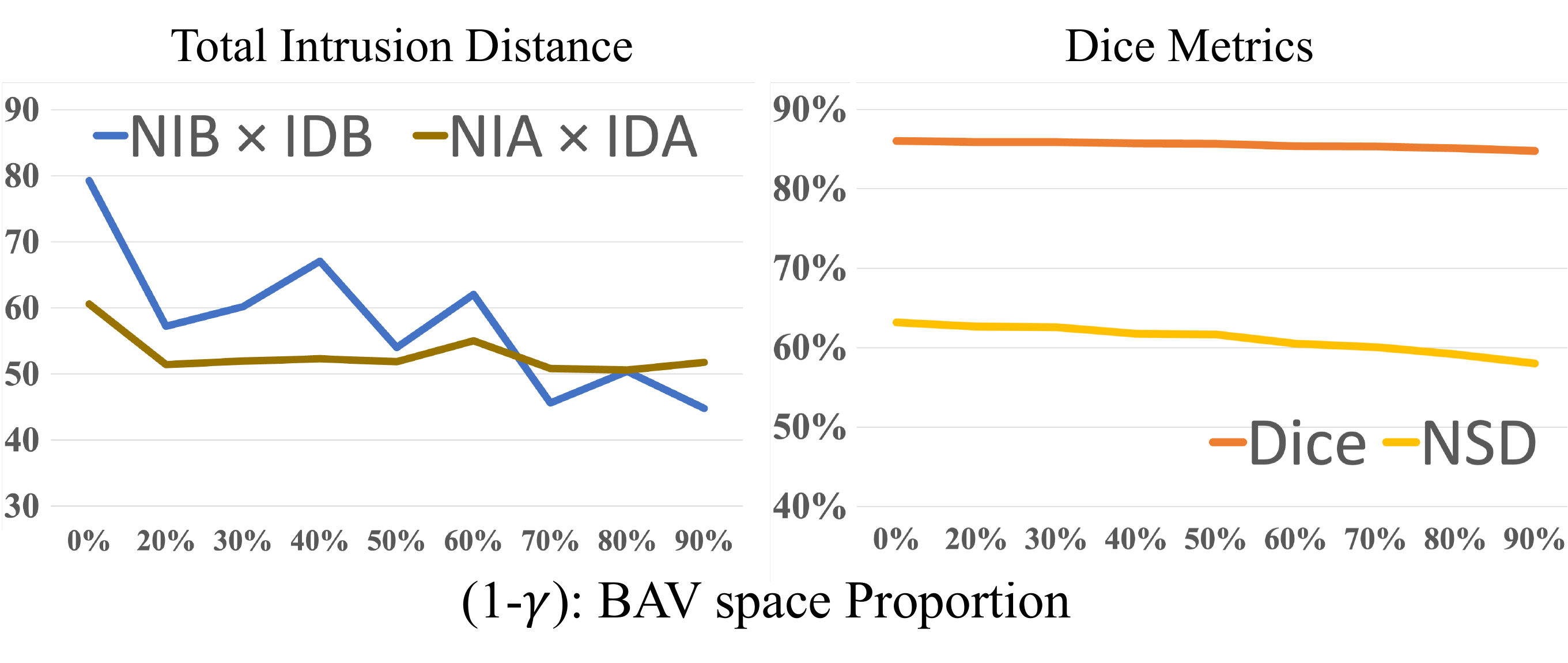}
	\caption{\textbf{Ablation on foreground point sampling in training.} The performance of \impulsep{ImPulSe+} with different proportions of foreground point ({\it{BAV}}) sampling, by total intrusion distance (Number of Intrusion ${\times}$ Intrusion Distance) and two dice metrics.  $1-\gamma$: the proportions of points that are sampled from points within the foreground{\it{BAV}} structures.} 
\label{fig:bav_sam_experiment}
\end{figure}

Fig. \ref{fig:bav_sam_experiment} shows all metrics of the \impulsep{ImPulSe+} network under various amounts of {\it{BAV}} sampling. As the proportion of {\it{BAV}} point sampling grows, the 2 dice metrics suffer from minor reduction while the total intrusion distance decreases tremendously. 
The results indicate that {\it{BAV}} sampling during the training stage mitigates the intrusion of bronchi and arteries, despite a detrimental influence on the dice metrics. At an intermediate range from about 20\% to 50\%, a significant drop in total intrusion distance is observed but the drawback in the dice metrics remains negligible implying that the {\it{BAV}} sampling strategy is overall beneficial to pulmonary segment segmentation, where anatomical-level quality is salient. Nevertheless, it is important to recognize that the elevated proportion of {\it{BAV}} sampling might precipitate over-fitting, potentially impinging upon the model's generalization capacity. 

\subsubsection{CNN Backbones}
\label{subsec:exp_backbone}
In this section, we evaluate the performance of the proposed architecture using different CNN networks as encoder $f$ (Fig. \ref{fig:pipeline}). As nnUNet~\citep{Isensee2021Nnunet} achieved noteworthy performance (Tab. \ref{tab:baselines}), we explore the impact of utilizing its backbone, 3D-UNet~\citep{iek20163DUL_3dunet}, as well as the 3D-UNet encoder, along with the backbone selection in \cite{Kuang2022WhatMF}, Resnet-18~\citep{He2015DeepRL_resnet}, as candidates for the CNN feature extractor $f$. 
For the 3D-UNet encoder and Resnet-18, the feature maps in each layer of the downsampling path are extracted and stacked as the input feature maps for the MLP layers of \impulsep{ImPulSe+}. For the entire 3D-UNet, the feature maps of each upsampling layer are utilized. The experiment results for the candidates are presented in Tab. \ref{tab:backbone_selection}.


The results show that in terms of Dice score, three backbone options achieve similar performance while ResNet-18~\citep{He2015DeepRL_resnet} is marginally better. \impulsep{ImPulSe+} with ResNet-18 or 3D-UNet as backbone acquires better NSD performance than 3D-UNet encoder by 1.5\% at minimum. For anatomical-level metrics, ResNet-18 records the best NIB and NIA performance. Therefore, considering the results across all metrics, applying ResNet-18~\citep{He2015DeepRL_resnet} as the backbone achieved the best.

\begin{table}
\centering
\caption{\textbf{Ablation on CNN backbones.}
The impact of various backbone architectures as CNN encoder $f$ for \textbf{\impulsep{ImPulSe+}} is presented.
}
\label{tab:backbones_compare}
\resizebox{\linewidth}{!}{
\begin{tabular}{lllllll}
\hline
CNN Encoder    & Dice (\%, $\uparrow$)    & NSD (\%, $\uparrow$)    & NIB ($\downarrow$)  & IDB ($\downarrow$)  & NIA ($\downarrow$)   & IDA ($\downarrow$)   \\ \hline
ResNet-18~\citep{He2015DeepRL_resnet}     & \textbf{86.00} & 63.17          & \textbf{24.09} & 3.29          & \textbf{42.39} & 1.43             \\
UNet~\citep{iek20163DUL_3dunet} Encoder & 85.42          & 61.68          & 24.96          & \textbf{2.41} & 43.87          & 1.43                \\
UNet~\citep{iek20163DUL_3dunet}   & 85.71          & \textbf{63.67} & 24.43          & 2.89          & 44.24          & \textbf{1.40}                    \\ \hline
\end{tabular}}
\label{tab:backbone_selection}
\end{table}

\begin{table}[ht]
\centering
\caption{\textbf{Performance of \impulsep{ImPulSe+} under different input combinations.}
Inputs include CT image (I), lobe (L), bronchi (B), artery (A), and vein (V).
Voxel-level metrics include Dice and NSD; anatomical-level metrics include intrusion-based measures (NIB, IDB, NIA, IDA).
Group labels indicate whether shapes are ground truth, predicted, or noisy predicted.}
\label{tab:input_combinations}
\resizebox{\linewidth}{!}{
\begin{tabular}{c|ccccc cc cccc}
\hline
& \multicolumn{5}{c}{\textbf{Input}} & \multicolumn{2}{c}{Voxel-level ($\uparrow$)} & \multicolumn{4}{c}{Anatomical-level ($\downarrow$)} \\
\cline{2-12}
\textbf{Group} & I & L & B & A & V & Dice (\%) & NSD (\%) & NIB & IDB & NIA & IDA \\
\hline
\shortstack{Image} 
& \checkmark & -- & -- & -- & -- & $85.87_{\pm .07}$ & $56.05_{\pm .06}$ & $23.98_{\pm .12}$ & $2.76_{\pm .22}$ & $49.67_{\pm .83}$ & $1.55_{\pm .03}$ \\ 
\hline
\multirow{5}{*}{\shortstack{GT \\ Shape}}
& \checkmark & -- & \checkmark & \checkmark & \checkmark & \textcolor{gold}{$86.48_{\pm .10}$} & \textcolor{gold}{$62.83_{\pm .07}$} & $24.21_{\pm .23}$ & \textcolor{silver}{$2.67_{\pm .16}$} & $43.29_{\pm .54}$ & \textcolor{gold}{$1.33_{\pm .12}$} \\ 
& -- & \checkmark & \checkmark & \checkmark & \checkmark & $82.56_{\pm .03}$ & $38.38_{\pm .10}$ & \textcolor{gold}{$20.65_{\pm .27}$} & $3.59_{\pm .47}$ & \textcolor{gold}{$38.18_{\pm .31}$} & $1.83_{\pm .04}$ \\ 
& \checkmark & -- & \checkmark & -- & -- & \textcolor{silver}{$85.52_{\pm .03}$} & \textcolor{silver}{$57.67_{\pm .06}$} & $22.86_{\pm .27}$ & \textcolor{gold}{$2.42_{\pm .08}$} & $48.68_{\pm .80}$ & \textcolor{silver}{$1.42_{\pm .03}$} \\ 
& -- & \checkmark & \checkmark & -- & -- & $79.50_{\pm .04}$ & $31.34_{\pm .12}$ & \textcolor{silver}{$20.83_{\pm .31}$} & $4.10_{\pm .25}$ & \textcolor{silver}{$40.60_{\pm .12}$} & $2.32_{\pm .11}$ \\ 
& -- & \checkmark & -- & -- & -- & $74.17_{\pm .18}$ & $25.53_{\pm .22}$ & $25.08_{\pm .40}$ & $7.40_{\pm .33}$ & $50.82_{\pm .23}$ & $3.75_{\pm .19}$ \\ 
\hline
\multirow{5}{*}{\shortstack{Pre-Seg \\ Shape}}
& \checkmark & -- & \checkmark & \checkmark & \checkmark & $86.06_{\pm .05}$ & $62.75_{\pm .07}$ & $23.66_{\pm .04}$ & $2.91_{\pm .11}$ & $47.07_{\pm .96}$ & $1.42_{\pm .13}$ \\ 
& -- & \checkmark & \checkmark & \checkmark & \checkmark & $82.43_{\pm .00}$ & $38.12_{\pm .08}$ & $21.22_{\pm .52}$ & $3.64_{\pm .11}$ & $39.92_{\pm .93}$ & $2.00_{\pm .04}$ \\ 
& \checkmark & -- & \checkmark & -- & -- & $85.55_{\pm .00}$ & $57.77_{\pm .06}$ & $23.61_{\pm .01}$ & $2.48_{\pm .21}$ & $48.46_{\pm 1.17}$ & $1.42_{\pm .01}$ \\ 
& -- & \checkmark & \checkmark & -- & -- & $79.46_{\pm .06}$ & $31.04_{\pm .09}$ & $21.29_{\pm .27}$ & $4.77_{\pm .38}$ & $41.14_{\pm .61}$ & $2.51_{\pm .15}$ \\ 
& -- & \checkmark & -- & -- & -- & $71.17_{\pm .24}$ & $22.39_{\pm .23}$ & $25.27_{\pm .14}$ & $7.87_{\pm .06}$ & $51.58_{\pm .40}$ & $3.96_{\pm .16}$ \\ 
\hline
\multirow{5}{*}{\shortstack{Noisy \\ Pre-Seg \\ Shape}}
& \checkmark & -- & \checkmark & \checkmark & \checkmark & $85.13_{\pm .10}$ & $56.33_{\pm .22}$ & $23.34_{\pm .08}$ & $2.61_{\pm .20}$ & $47.50_{\pm .52}$ & $1.34_{\pm .08}$ \\ 
& -- & \checkmark & \checkmark & \checkmark & \checkmark & $74.42_{\pm 1.41}$ & $28.51_{\pm .88}$ & $21.06_{\pm .07}$ & $2.27_{\pm .04}$ & $42.81_{\pm .48}$ & $1.20_{\pm .06}$ \\ 
& \checkmark & -- & \checkmark & -- & -- & $85.39_{\pm .05}$ & $57.29_{\pm .17}$ & $23.50_{\pm .14}$ & $2.31_{\pm .08}$ & $49.04_{\pm .59}$ & $1.44_{\pm .06}$ \\ 
& -- & \checkmark & \checkmark & -- & -- & $76.49_{\pm .16}$ & $28.06_{\pm .13}$ & $22.28_{\pm .09}$ & $3.87_{\pm .24}$ & $47.90_{\pm .52}$ & $2.21_{\pm .04}$ \\ 
& -- & \checkmark & -- & -- & -- & $71.17_{\pm .24}$ & $22.39_{\pm .23}$ & $25.27_{\pm .14}$ & $7.87_{\pm .06}$ & $51.58_{\pm .40}$ & $3.96_{\pm .16}$ \\ 
\hline
\end{tabular}}
\label{tab:input_ablation}
\end{table}

\section{\revisedtwo{Shape-based Reconstruction: From Shapes to Segments}}
\label{sec:shape}

\subsection{\revisedtwo{Experiment Setting}}
\revisedtwo{In contrast to Section 5, which examined reconstruction performance driven primarily by CT image inputs, we now focus on shape-based reconstruction. In this setting, we experiment with reconstruction by anatomical structures rather than by image intensities. Shape-based reconstruction is clinically meaningful because these structures directly determine segment borders~(Fig.~\ref{fig:pulmonary_segment_anatomy}. D) and provide strong priors for anatomically consistent outcomes~(Sec.~\ref{subsec:ps_reconstruction}). It is also practical since shapes data are relatively free from privacy concerns compared to images. In this section, we study the effect of using shape as primary input~(Sec.~\ref{subsec:shape_contribution}), performance under out-of-distribution shape input~(Sec.~\ref{subsec:shape_OOD}), and performance when only partial shape input is available~(Sec.~\ref{subsec:shape_partial}). All subsequent shape-related analyses in this section are summarized in Tab.~\ref{tab:input_ablation}, where the input combination is indicated by the check-mark, each additional experiment was repeated three times.}

\subsection{\revisedtwo{Contribution of Shape Inputs}}
\label{subsec:shape_contribution}
\revisedtwo{In this section, the contribution of anatomical shapes to pulmonary segment reconstruction is examined and analyzed. According to Tab.~\ref{tab:input_ablation}, using only CT images yielded satisfactory voxel-level performance, showing that intensity information alone could provide globally accurate segments. When image is combined with the ground-truth {\it BAV} shapes, voxel-level performances improve trivially~(.7\%) but improves dramatically across the anatomical-level metrics~(1\%, 3.3\%, 12.8\%, 14.2\%). This improvement shows that while shape information is already embedded in the image, explicitly pre-delineating it and adding it to the input provides additional guidance, helping the network focus on these structures and enforce anatomical correctness.

Interestingly, if CT image input in the {\it IBAV} setting is replaced by the lobe shape and becomes {\it LBAV}, making this a shape-only input combination, we observe a sharp voxel-level degradation of 4.5\% Dice and 38\% NSD. However, this setting produces fewer intrusions across bronchi~(14.7\%) and arteries~(11.8\%), suggesting that structural priors alone can enforce anatomical consistency even when voxel-level accuracy is reduced.

Overall, these findings show the complementary roles of image and anatomical structures as well as showcasing the strong anatomical prior that shape-based input brings. While CT images provide strong voxel-level accuracy, structural inputs enforce anatomical plausibility and reduce clinically relevant intrusions.}

\subsection{\revisedtwo{Out-of-distribution Shape Evaluation}}
\label{subsec:shape_OOD}
\begin{figure}
    \centering
    \includegraphics[width=1\linewidth]{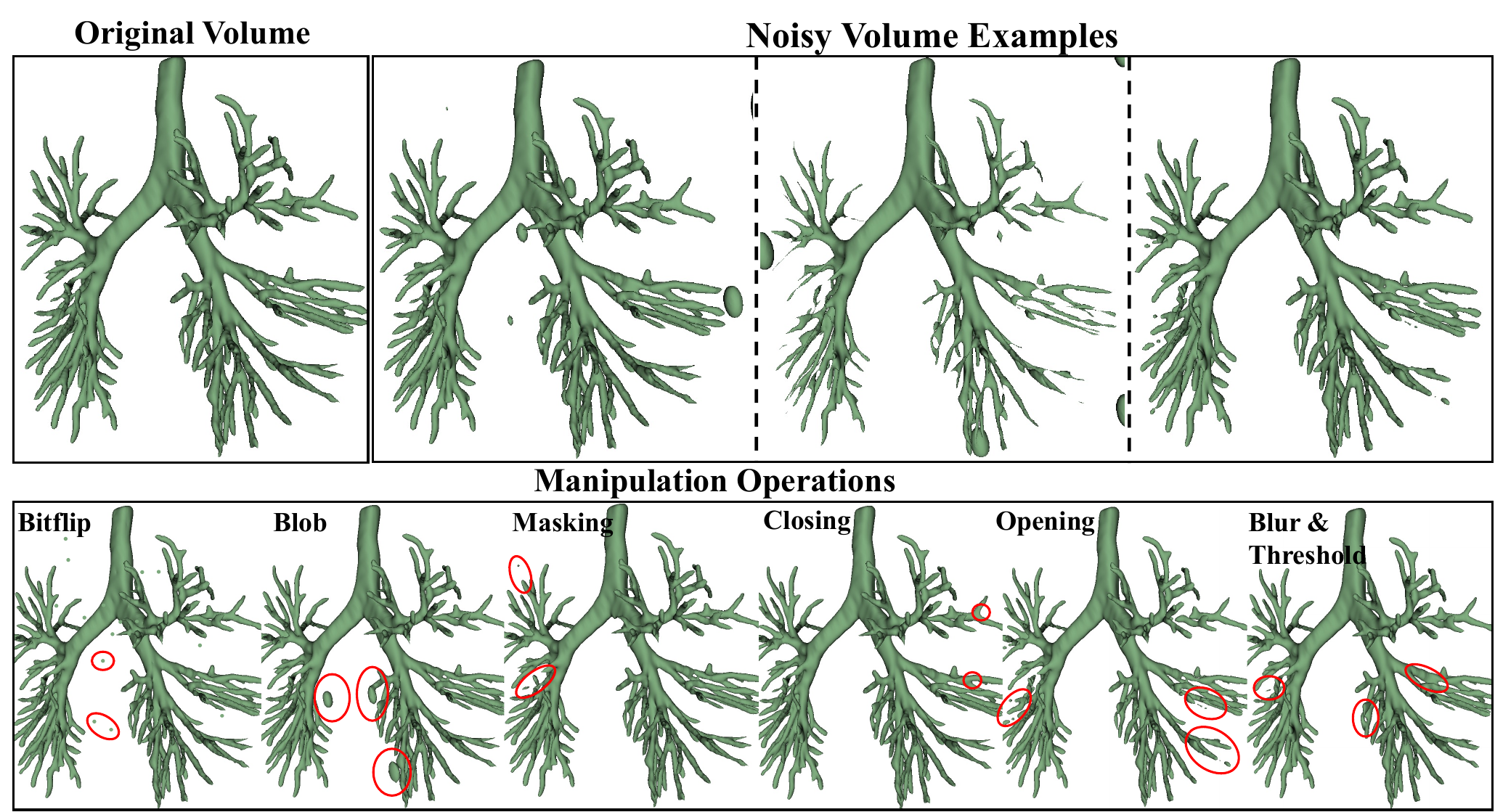}
    \caption{The top images show examples of the noisy volumes after random manipulation operations compared against the original volume. The bottom volumes are illustrations of the effect of each individual manipulation operations.}
    \label{fig:noisy_shape}
\end{figure}

\revisedtwo{As outlined in Sec.~\ref{subsec:exp_setting}, we train a separate segmentation network to acquire predicted {\it BAV} shapes as shape priors and concatenate them with image inputs. These predicted shapes now serve as a convenient benchmark for assessing the network’s capability when operating with non–ground-truth shape-based inputs, referred to as Pre-Seg shapes, also presented in Tab.~\ref{tab:input_combinations}.

Focusing on the {\it LBAV} input setting, we observe trivial voxel-level performance difference between GT and Pre-Seg shapes. The performance degradation with Pre-Seg shape is particularly pronounced in anatomical-level due to the loss of ground-truth information. However, it still delivers competitive results, better than the image-only counterpart in the number of intrusion~(NI~A/B) aspect. This suggests that automatically derived shapes can capture much of the structural information necessary for reconstruction, though high-quality GT inputs remain essential for achieving the highest level of anatomical fidelity.

To further stress-test the model, we generate noisy variants of the Pre-Seg shapes through randomized manipulations, referred to as "Noisy Pre-Seg shape" in Tab.~\ref{tab:input_combinations}. Each manipulation has a 30\% probability of being applied and the application order is randomized. Shown in Fig.~\ref{fig:noisy_shape}, the manipulations include voxel flipping, morphological opening and closing, insertion or masking using random ellipsoid blobs, and Gaussian filtering followed by thresholding. These manipulations produce corrupted yet anatomically plausible shapes. When compared against the ground-truth, the Noisy Pre-Seg shapes has average dice scores of 64.2\%, 75.9\%, 75.2\% for {\it B, A, V} respectively. Evaluating reconstruction on both Pre-Seg and Noisy Pre-Seg shapes provides a controlled framework for studying out-of-distribution robustness in shape-based reconstruction.

With noisy inputs, we observed that voxel-level performance tend to degrade significantly when the image modality was not included~({\it LBAV, LB, L}), reflecting that voxel-level accuracy relies heavily on image information. In contrast, anatomical-level metrics remained largely stable, even though the manipulated shapes could differ substantially from the original inputs. This indicates that anatomical fidelity depends less on the image and more on structural cues. The reason is that, despite shape manipulations, the trajectory and spatial location of the tree-like structures were preserved, and these implicitly define intersegmental borders, enabling the model to maintain consistent anatomical correctness.}
\subsection{\revisedtwo{Sensitivity to Partial Shape Input}}
\label{subsec:shape_partial}
\revisedtwo{Beyond robustness to distribution shifts, it is necessary to examine how performance changes when only partial shape information is available. In practice, bronchial structures are more consistently segmented, whereas arteries and veins are less reliably available due to their smaller size, greater anatomical variability, and lower contrast in CT images. This motivates experiments with progressively reduced structural input. 

{\it LBAV}, our default for shape-based reconstruction, discards image intensity and relies exclusively on shapes, testing the capacity of geometric priors alone. IB retains images and bronchi but omits arteries and veins, capturing a frequent real-world scenario where vascular information is missing. Further reductions to LB (lobes and bronchi) or L (lobes only) simulate minimal inputs that may still provide coarse anatomical guidance. Each of these settings is evaluated with ground-truth shapes, pre-segmented shapes, and noisy pre-segmented shapes, providing a comprehensive assessment of sensitivity to partial input.

In both the ground-truth and predicted shape scenarios, removing arteries and veins ({\it A,V}) from the full {\it LBAV} input led to a clear drop in voxel-level metrics and a substantial degradation in intrusion distance measures. The dramatic increase in average intrusion distance can be explained by the fact that arteries and veins span large regions within their respective segments and serve as key references for boundary placement. When these structures are absent, the model loses explicit spatial anchors to delineate intersegmental planes, and reconstructed segments are more likely to intrude into adjacent regions, thereby increasing the average distance of boundary violations. With only lobes and bronchi provided, such guidance is significantly reduced. Furthermore, when bronchi are also removed, leaving only lobar information, both voxel-level and anatomical-level performance degraded further, underscoring the critical role of bronchi and vessels in constraining the reconstruction to anatomically valid boundaries.}

\section{Conclusion}



This paper presents \impulsep{ImPulSe+} to automate the reconstruction of pulmonary segments anatomy using neural implicit functions. By introducing a Deformation network and a Correction network in cooperation with the learned template, the \impulsep{ImPulSe+} network achieves superior segmentation outcomes and surpasses all counterparts in both DSC and NSD metrics. Besides, we propose two new metrics for anatomy-level evaluation considering the anatomical constraint in pulmonary segments. Finally, we developed the \emph{Lung3D} dataset (Fig.~\ref{fig:dataset}), which is the first open dataset for pulmonary segment segmentation. Within this dataset, we investigate the feasibility of shape-based fine-grained pulmonary segment reconstruction, and our proposed methodology demonstrates encouraging outcomes. 

\revisedtwo{Beyond the primary evaluations, we compared our method to SOTA methods and thoroughly test the method under noisy and partial input shapes. We further carry out a sensitivity analysis on rare anatomical variations and a failure-case study to examine how the deformation and correction networks behave under difficult scenarios. These analyses highlight the robustness of ImPulSe+ in handling challenging anatomical variations.}

In future works, we shall explore incorporating intersegmental veins as border guidelines and leveraging diffeomorphism to explicitly preserve the topology of reconstructed pulmonary segments. Moreover, it is also interesting to investigate how to reconstruct pulmonary segments in more challenging cases with disconnected pulmonary structures~\citep{weng2023topology}. 

\section*{Acknowledgements}
J.Y was supported by the ELLIS Institute Finland and School of Electrical Engineering, Aalto University. M.G. was supported by the US NSF CAREER award IIS-2239537. This work was supported in part by the Swiss National Science Foundation. 

\section*{Data and Code Availability}
The \emph{Lung3D} dataset, together with the \impulsep{ImPulSe+} code, will be available at \url{https://github.com/HINTLab/ImPulSe}.

\bibliographystyle{model2-names.bst}\biboptions{authoryear}
\bibliography{strings,refs}

\end{document}